\pdfoutput=1
\documentclass[prd,twocolumn,english,nofootinbib]{revtex4-1}
\usepackage{natbib}
\usepackage{bm}
\usepackage{epsfig}

\usepackage[T1]{fontenc}
\usepackage[latin9]{inputenc}
\usepackage{amsmath}
\usepackage{amssymb}
\usepackage{graphicx}
\usepackage{esint}

\def\mnras{M.N.R.A.S}
\def\apj{Astrophys.J.}
\def\apjl{Astrophys.J.Lett.}
\def\apjs{Astrophys.J.Supp.}

\def\prd{Phys.Rev.D}
\def\araa{Annu.Rev.Astron.Astrophys.}
\def\aap{Astron.Astrophys.}

\def\Mesz{M\'esz\'aros~}

\begin{document}
\title{Multi-GeV Neutrino Emission from \\ Magnetized Gamma Ray Bursts}
\author{Shan Gao and Peter \Mesz}
\affiliation{Department of Physics, Department of Astronomy and Astrophysics,\\
Center for Particle Astrophysics,\\
The Pennsylvania State University, University Park, 16802, USA}
\date{\today}

\begin{abstract}
We investigate the expected neutrino emissivity from nuclear collisions in a
magnetically dominated model of gamma-ray bursts motivated by recent
observational and theoretical developments. The results indicate that small
multi-GeV neutrino fluxes are expected for model parameter values which are
typical of electromagnetically detected bursts. We show that for detecting at
least one muon event in Icecube and its Deep Core sub-array, a single burst
must be near the high end of the luminosity function, and at low redshifts
$z\lesssim 0.1$, where the burst rate is very low. We also calculate the
luminosity and distance ranges that can generate $0.01-1$ muon events per GRB
in the same detectors, which may be of interest if simultaneously detected
electromagnetically, or if measured with future extensions of Icecube or other
neutrino detectors with larger effective volume and better sensitivity.
\end{abstract} \maketitle

\section{Introduction}
\label{sec:intro}

Observations of multi-GeV photons from gamma-ray bursts (GRB) recently 
accumulated by the {\it Fermi} satellite (e.g. \cite{Abdo+09grb080916c,
Ackermann+10grb090510}) have pointed out the need to re-evaluate the type of
models used to explain the prompt photon emission mechanisms and the location
of the emission regions in these objects (e.g. \cite{Peer11-fermigrb}).  In
particular, concerns about the radiative efficiency of usual internal shock
models, and the larger radii required to avoid two-photon degradation of the
spectra have spurred the investigation of baryonic (non-MHD) jet models where
the radiation arises in a jet photosphere \citep{Meszaros+00phot,
Rees+05photdis,Beloborodov10phot}. In such baryon-loaded jet models, at a
certain radius the timescale of the nuclear elastic collisions (which couple
the proton $p$ and neutron $n$ components) becomes longer than the expansion
timescale, i.e. the collision optical depth falls below $o(1)$, and the two
components decouple from each other \citep{Derishev+99,Bahcall+00pn}. The
protons can continue to be accelerated by the radiation, while the neutrons,
which have zero electric charge, start to coast with a constant Lorentz factor.
Starting at this decoupling radius and for some distance beyond, the
longitudinal drift velocity between the $n$ and $p$ becomes $\Delta v \gtrsim
0.5 c$, and they start to collide inelastically.  Such large relative
velocities between $n$ and $p$ components also can arise in realistic jets
where the bulk Lorentz factor $\Gamma$ depends on the polar angle $\theta$,
which also leads to inelastic collisions \citep{Meszaros+00gevnu,
Beloborodov10phot}, as neutrons from the outer parts (sheath) of the jet
thermally drift into the jet core. In both pictures, pions are created, which
in the case of the dynamics being dominated by the baryons results in multi-GeV
photons and neutrino production \citep{Bahcall+00pn,Beloborodov10phot}. 

A different approach towards resolving the radiative efficiency of GRBs
involves consideration of magnetically dominated jets
\citep{Giannios+07photspec, Tchekhovskoy+10grb,Meszaros+10pop3}. Some of these
magnetic models assume an almost baryon-free outflow
\citep{Usov94,Meszaros+97poynting,Meszaros+10pop3, Toma+11pop3,Gao+11pop3nu},
while in other cases a  substantial but dynamically sub-dominant baryon load is
assumed \citep{Thompson94,Drenkhahn+02,Lyutikov+03grbmag,
Giannios+07photspec,Tchekhovskoy+10grb,McKinney+11magphot,Metzger+11grbmag,Giannios07pn}.
The baryons in such jet models are expected to accelerate at a different rate
than in baryonic (non-MHD) jet models, and the different dynamics leads to
quantitatively different predictions for the photon spectrum
\citep{Meszaros+11gevmag}. Similarly, it should lead to quantitatively
different neutrino spectra, which we investigate in this paper.

Unlike in the previous investigations cited above, here we consider the
neutrino spectra arising from nuclear collision effects in magnetically
dominated GRB outflows.  In this case, both the radial $np$ decoupling  radius
as well as the photosphere occur at larger radii from the central engine  than
in the non-magnetic case, and also the transverse drift of neutrons from the
periphery of the jet into the jet core becomes important at different radii,
where the physical conditions differ from those previously considered. As a
consequence, multi-GeV photons are produced at somewhat softer energies and
with appreciable time delays \citep{Meszaros+11gevmag} respect to the MeV
photospheric photons, but the detailed neutrino spectrum for such magnetically
dominated jets has not been considered so far. In this paper we investigate
numerically the neutrino spectrum expected in magnetically dominated
baryon-collisional GRB models, taking into account both longitudinal and
transverse $n,p$ decoupling and inelastic collisions.  These neutrinos are in
the energy sensitivity range of Icecube and its DeepCore
\citep{DeYoung+11RICAP,IcecubeDC+09,IcecubeICRC+11} sub-array.

In \S \ref{sec:col} we briefly introduce the astrophysical model and present
the method of neutrino emission calculation. In \S \ref{sec:nuflux} we present
the results for the expected neutrino fluxes and muon events, as well as the
detection prospects with Deep Core and IceCube. A discussion and summary of the
results is given in \S \ref{sec:disc}.

\section{The Nuclear Collision Scenarios}
\label{sec:col}

We consider two types of nuclear collision scenarios, one where the collisions
occur as a result of longitudinal (radial) velocity drifts, and another where
they occur as result of transverse (relative to the jet axis) drifts of
neutrons from an outer jet sheath to an inner jet core where the bulk Lorentz
factor is different.  We consider both of these cases in the context of
magnetically dominated jet dynamics, which differs from the usually considered
baryonically dominated dynamics.

\subsection{Neutrinos From Longitudinal Nuclear Collisions}
\label{sec:long}

In a magnetized outflow the bulk Lorentz factor accelerates initially as
\citep{Drenkhahn02,Granot11bmagjet,Metzger+11magcr}
\begin{equation}
\Gamma=\begin{cases}
(r/r_{0})^{1/3} & r<r_{sat}
\label{eq:Gamma}
\\
\eta & r>r_{sat}
\end{cases}
\end{equation}
where $r_{sat}=\eta^{3}r_{0}$ is the saturation radius beyond which the jet
material starts to coast. (This is in contrast to the baryonic dominated
dynamics, where $\Gamma \propto r$ up to an $r_{sat}\sim \eta r_0$).

At the radius where the co-moving baryon collision timescale becomes longer
than the adiabatic expansion timescale, the protons decouple from the neutrons,
and they continue to accelerate as $\Gamma\propto r^{1/3}$. The condition above
is expressed as \begin{equation}
t_{exp}^{\prime}=r/c\Gamma<t_{col}^{\prime}=1/n_{b}^{\prime}\sigma_{\pi}c
\label{eq:tdec} \end{equation} where \begin{equation}
n_{b}^{\prime}=L_{tot}/4\pi r^{2}c\eta\Gamma m_{p}c^{2} \label{eq:nbar}
\end{equation} is the total comoving baryon density in the flow, $L_{tot}$ is
the jet total isotropic equivalent  luminosity, and $\eta=L_{tot}/{\dot M c^2}$
is the dimensionless entropy or energy to mass outflow ratio.

The neutrons thereafter coast with a bulk Lorentz factor
$\Gamma_{n}=(r_{d}/r_{0})^{1/3}$, where 

\begin{equation} r_{d}=\eta_{\pi}^{3}(\eta_{\pi}/\eta)^{3/5}r_{0}
\end{equation} is the decoupling radius $r_{d}\approx r_{\pi}$ defined from the
condition (\ref{eq:tdec}) (see also MR11). Here $\eta_{\pi}$ is a dimensionless
parameter given by

\begin{equation} \eta_{\pi}=(\frac{L_{tot}\sigma_{\pi}}{4\pi
cm_{p}c^{2}r_{0}})^{1/6}\approx1.33\times10^{2}L^{1/6}_{54}r^{-1/6}_{0,7}
\end{equation} where $\sigma_{\pi}\sim3\times10^{-26}cm^{-2}$. Decoupling
occurs at $r_d < r_{sat}$ if $\eta >\eta_\pi$, otherwise, if the condition
(\ref{eq:tdec}) is met at a radius above $r_{sat}$, decoupling never happens
and the $p$ and $n$ just coast together.
 
Beyond the decoupling radius, the accelerated protons collide longitudinally
\footnote{neglecting random thermal motions and transverse collisions, which
are discussed in the next section.} with the neutrons, which have a smaller
Lorentz factor. The collisions becomes mostly inelastic when their relative
Lorentz factor $\Gamma_{rel}\gtrsim1.3$; see eq.  (\ref{eq:Gammarel}).  In the
star frame , each shell encompassed within $(r,r+dr)$ contributes to the
pionization optical depth for those protons by an amount \begin{equation}
d\tau(r)=n_{n}(r)\Delta\beta\sigma_{\pi}dr \label{eq:tau} \end{equation} where
$\Delta\beta$ is the relative speed between protons and neutrons in the
lab-frame, $\Delta\beta=[1-\Gamma^{-2}(r)]^{1/2}-[1-\Gamma^{-2}(r_{d})]^{1/2}$

To estimate the pion spectrum, we can, for simplicity, assume that for each
collision, in the center of mass frame the proton and neutron are approximately
at rest after the collision, with a maximum number of pions created, which are
also approximately at rest (we will use a more detailed treatment in \S
\ref{sec:trans}).  Therefore, the invariant energy  is \begin{equation}
\sqrt{s}=\sqrt{(p_{p}^{\mu}+p_{n}^{\mu})^{2}}=\sqrt{2(1+\Gamma_{rel})}
\label{eq:sqrts} \end{equation} where the proton Lorentz factor viewed in
neutron co-moving frame is \begin{equation}
\Gamma_{rel}=\frac{1}{2}(\frac{\Gamma_{p}}{\Gamma_{n}}+\frac{\Gamma_{n}}{\Gamma_{p}})
\label{eq:Gammarel} \end{equation} This relation is valid when both
$\Gamma_{p},\Gamma_{n}\gg1$ .

The maximum number of pions that can be created (the pion multiplicity) is
\begin{equation} \lambda_{\pi}=(\sqrt{2(1+\Gamma_{rel})}-2)/(m_{\pi}/m_{p})
\label{eq:lambdapi} \end{equation} with a Lorentz factor \begin{equation}
\Gamma_{\pi}(r)=\frac{\Gamma_{p}(r)+\Gamma_{d}}{\sqrt{2[1+\Gamma_{rel}(r)]}}
\label{eq:Gammapi} \end{equation} The radius-dependent probability for a proton
to interact with a neutron is \begin{equation} P(r)=e^{-\tau(r)} \end{equation}
where $\tau(r)=\int_{r_{d}}^{r}d\tau$ and $d\tau(r)$ is given by eqn.
(\ref{eq:tau}) The number of pions created per proton is \begin{equation}
N_{\pi}=\int_{r_{d}}^{r_{max}}\lambda_{\pi}dP(r) \end{equation} where the
maximum interaction radius is estimated as $r_{max}\sim
ct_{duration}/\Delta\beta\sim3r_{sat}$ (the last equality is for the choice of
nominal parameters in MR11. Otherwise $r_{max}\sim ct_{n-decay}\Gamma_{n}$).

\begin{figure} \includegraphics[width=0.7\columnwidth]{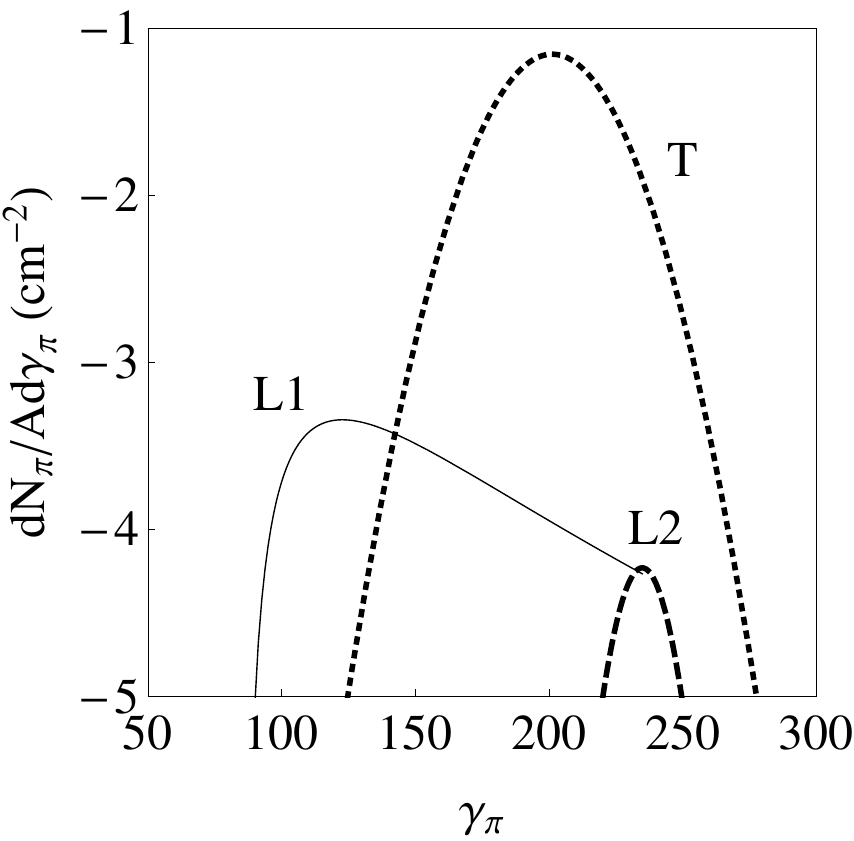} \caption{
Charged pion spectrum at the source from longitudinal and transverse nuclear
collisions (analytical approximation), normalized to the Earth observer frame.
The charged pions decay in the source, and are observed at Earth only via their
neutrino decay products.  The component L1 is from longitudinal $pn$ collisions
at radii $r_{d}<r<r_{sat}$, while the component L2 arises at $r>r_{sat}$. The
component T is from transverse drift $pn$ collisions (see \S \ref{sec:trans}).
A Gaussian dispersion is assumed for simplicity to represent the broadening of
the real spectrum arising from various effects (see \S \ref{sec:long}). The
dominant pion (and neutrino) production comes from the transverse $pn$
collisions, as discussed in \S \ref{sec:trans}.  We assume a source with
$L_{\gamma}=0.1L_{tot}=10^{53}erg/s$, $\eta=500$, $\theta_{jet}=0.01$ and
$z=0.1$ (see \S \ref{sec:param}).} \label{pionLT}   \end{figure}

Beyond the decoupling radius $r_{d}$, the protons continue to be accelerated
until reaching $r_{sat}$. When the energy is above the threshold to create one
pion in a collision with a neutron, we define this radius as, e.g. $r_{1}$ . At
a larger radius, protons with greater energies create more pions per collision.
The pion Lorentz factor as a function of their production radius is given by
eqn. (\ref{eq:Gammapi}). The inelastic interactions starts from $r_{1}$ and may
last to a radius above $r_{sat}$. Beyond $r_{sat}$, the protons coast with a
Lorentz factor $\eta$ and (provided $\eta>\eta_\pi$) create pions with a
monochromatic energy (in the first-order approximation used in this section).
In fact, the resulting neutrino spectrum, after the pions decay, will be
broadened due to various factors, e.g. a) thermal motions of both protons and
neutrons; b) energy dispersion of the created charged pions; c) kinematics of
the pion and muon decay process; d) pion and muon cooling and re-scattering.
In this \S \ref{sec:long} the above four factors are not considered in the
calculation, except in a qualitative way. This is adequate because, as
discussed in \S \ref{sec:trans} and in Fig.[\ref{pionLT}], it turns out that the dominant
process for neutrino production is through transverse nuclear collisions.  As
an example, for a case with $\eta=500$ , $r_{0}=r_{0,7}$ , $L_{tot}=L_{54}$ ,
$L_{p}=L_{n}=0.5L_{b}$ (where the sub-index $p,n,b$ refer to proton, neutron,
baryon luminosity respectively), the charged pion spectrum from longitudinal
nuclear collision is shown in Fig.[\ref{pionLT}].  For simplicity reasons, a
Gaussian distribution is assumed to represent the dispersion.  The pion
spectrum from transverse collisions is also shown in the figure, anticipating
the results of the more detailed treatment of the latter in \S \ref{sec:trans}
and after.

\bigskip
\subsection{Neutrinos From Transverse Nuclear Collisions}
\label{sec:trans}

\bigskip

In  more realistic jet models, the jet properties vary in the transverse
($\theta$) direction. Hydrodynamical simulations indicate a smaller Lorentz
factor in the jet outer regions (jet edge) than that in the jet core
\cite{Zhang+03jetnum}, and qualitatively similar results appear also in some
MHD outflows \cite{Tchekhovskoy+08grb}.  As a simple ideal model, we consider
the transverse structure of the jet in the region outside the star to be
represented by a two-step function, consisting of an inner jet core with
$\Gamma$ give by eqn. (\ref{eq:Gamma}) and a slower outer jet, or jet sheath,
with a saturation Lorentz factor of $\eta_{out}=10^{2}\eta_{out,2}$. Both inner
and outer jets will have been populated with protons and free neutrons already
near the black hole, where any nuclei present would have been
photo-dissociated. Due to thermal diffusion, neutrons in the jet sheath can
drift sideways into the jet core and interact with baryons in the core. For
significant effects, this requires the neutrons in the jet sheath to drift
through a substantial transverse distance $\sim r\theta$ into the core. This
condition can be  roughly estimated as

\begin{equation}
r_{\perp}\gtrsim\eta_{\pi}^{6}\theta<\Gamma_{rel}>r_{0}/\eta, \label{eq:rperp}
\end{equation} 

where $<\Gamma_{rel}>$ is the average relative Lorentz factor
between the neutron and the baryons encountered along its path \footnote{The
path calculation is a well-defined but complicated problem. To derive eqn.
(\ref{eq:rperp}) we set the pionization optical depth $\tau_{\pi}=1$ along the
neutron path for which its transverse displacement amounts to $r\theta$. The
path itself is not transverse to the jet axis, due to relativistic beaming
according to different sheath and core bulk velocities. $<\Gamma_{rel}>$
depends on the details of the jet transverse structure, and the density along
the neutron path varies due to the jet dynamics. However, eqn. (\ref{eq:rperp})
serves as a rough estimate.}.  We estimate the number of neutrons which have
drifted into the core in a timescale $t_{\perp}$ (all measured in the star
frame)  as \begin{equation} N_{\perp}(t_{\perp})\sim\pi
r^{2}\theta\int_{0}^{t_{\perp}}\phi_{n}(t)dt \label{eq:Nperp} \end{equation}
where the diffusive flux $\phi_{n}(t)$ is from eqn. (14) in MR11. Meanwhile,
the number of baryons ($n$ and $p$) passing longitudinally through the jet core
is

\begin{equation} N_{\parallel}(t_{\perp})\sim\pi r^{2}\theta n_{b}ct
\label{eq:Npara} \end{equation} 

The number of collisions \emph{per baryon} in
the core is roughly $N_{\perp}/N_{\parallel}$, because the pionization optical
depth is $\tau\sim1$ for those neutrons. (The case
$N_{\perp}/N_{\parallel}\gg1$ is less likely for typical jet parameters, and
would involve protons in the core undergoing multiple scatterings resulting in
cascades, which is beyond the scope of this paper).  From the equations above,
\begin{equation} N_{\perp}/N_{\parallel}\sim
2\eta_{out}^{-1}\eta_{\pi}^{-3}(1-T^{\prime-2})^{1/2}(y\eta/t_{\perp})^{1/2}r_{\perp}
\label{eq:NN} \end{equation} 

where $r_{\perp}$ is expressed in units of $r_{0}$
and $T^{\prime}$ is the comoving temperature in units of $m_{p}c^{2}$
($T'\approx o(1)$ in this case, see MR11).  The parameter $y$ is the ratio of
neutron density in the sheath to baryon density in the core, $y\equiv
n_{n,out}/n_{b}$, in the star frame.

\subsection{Parameters of the Model}
\label{sec:param}

The neutrino spectrum at Earth depends on a number of parameters. Among these
are the total luminosity $L_{tot}$ of the jet, which is initially mostly in
magnetic form; the ``baryon" luminosity $L_{b}$, which is the the dominant
energy form beyond the saturation radius $r_{sat}=\eta^{3}r_{0}$ (the ratio of
these two being defined as $\epsilon_b=L_{b}/L_{tot}$); the photon luminosity
produced by dissipation process around the photosphere $L_{ph}$ (whose ratio to
the total initial luminosity is defined as $\epsilon_{ph}=L_{ph}/L_{tot}$).
In the inner jet core, the proton and neutron luminosity are assumed, for
simplicity, to be $L_{p}=L_{n}= (1/2)L_{b}$ throughout the paper \footnote{The
outer jet neutrons can also collide with jet core neutrons and produce
neutrinos. However, the relative Lorentz factor with neutrons is smaller than
that with the accelerated protons in the jet core, so the neutrino production
through $nn$ collisions is less efficient than for $pn$ collisions.  Only the
latter are discussed in this paper.}.  We take for the ratio $y$ of baryon
density in the outer jet to that in the inner jet a nominal value $y=0.01$.
The discussions on results from different $\eta_{core}$ and $\eta_{out}$ 
are presented in Figure [\ref{L},\ref{eta}].
We adopt a nominal jet opening angle of $\theta=0.01$, other
values being discussed in Figure [\ref{theta}]. We also adopt as a standard
burst duration in the source frame  a value of $t=20 s$, which is a rough
average value for long bursts.  The redshift-distance relation used is that for
a standard $\Lambda CDM$ cosmology.

\section{GeV Neutrinos and their Detectability in Deep Core and ICECUBE}
\label{sec:nuflux}

\subsection{Neutrinos from a single GRB}
\label{sec:nusingle}

The neutrinos are produced by nuclear collisions between protons and neutrons
leading to pions, the charged pions subsequently decaying \footnote{The charged
Kaon leading decay channel is similar and can be approximated as effective
pions. However, the number ratio of Kaons to pions produced by hadron collision
at these energies is less than 0.1 and can be neglected for the purpose of this
paper.} as
\begin{equation}
\pi^{\pm}\rightarrow\mu^{\pm}+\nu_{\mu}(\bar{\nu}_{\mu})\rightarrow
e^{\pm}+\nu_{e}(\bar{\nu}_{e})+\nu_{\mu}+\bar{\nu}_{\mu}~.  \label{eq:piontonu}
\end{equation}
The neutrino flavor mix produced at the source is determined by
eqn.[\ref{eq:piontonu}], but as a result of neutrino oscillations, the neutrino
flavor received at Earth depends on energy and distance. Since in this paper we
discuss the general case, not a specific GRB, we approximate the received
neutrino flux as having equal numbers in all three flavors. 

To calculate the neutrino spectrum from nuclear collisions the use of a
numerical code is desirable. The commonly used PHYTIA-8 code uses a minimum
threshold energy of $E_{cm}=10$ GeV, which for the energies considered here
leads to inaccuracies.  For this reason, we have used two different numerical
methods. One method uses the publicly available code of \cite{Kamae06code}.
These authors use a parametrization formula for $\gamma$,$e^{\pm}$,$\nu$,and
$\bar{\nu}$ which is carried out separately for  diffractive, non-diffractive
processes and resonance-excitation processes, with a logarithmic rising $pp$
inelastic cross section with $T_{p}$. The secondary particle spectra are
initially extracted out of events generated for monoenergetic protons
($0.488GeV<T_{p}<512TeV$), using several simulation programs. The spectra are
then fitted by a common parametrized function, separately for the physical
processes listed above. Finally, the parameters determined for monoenergetic
protons are fitted as a function of proton energy. The procedure was repeated
for all those secondary particle types mentioned above. In order to approximate
better the experimental data at lower energies, two baryon resonance
contributions have been included, one representing the $\Delta(1232)$ and the
other representing multiple resonances around 1600 $MeV/c^{-2}$.  However, as
pointed out also by \cite{Kamae06code}, the pion mean energy in this code is
slightly under-estimated at incident proton kinetic energy (in the fixed target
lab frame) of $T_{p}\sim2$ GeV and above, compared to  experimental data.
Therefore in this paper we have corrected this discrepancy by multiplying the
resultant neutrino energy by a factor of 1.3 using an approximate fit to the
Fig. 5 of their paper.  This causes a slightly over-estimated pion mean energy
near the threshold energy $T_{p}\sim0.3$ GeV; however, this energy range is not
of interest for our purposes in this paper, since it results in a
non-detectable neutrino flux is associated at the associated energy.  We refer
to this numerical calculation as method A. 

We have also developed a different code, which is independent of method A, in
order to cross-check the validity of method A around energies $T_{p}\gtrsim4$
GeV, and in order to gain better transparency on the underlying physical
processes where this is not otherwise made explicit in method A. We refer to
this second method as method B.  The energy $T_{p}\gtrsim 4$ GeV is of
particular importance here for at least three reasons: i) it arises naturally
in the astrophysical model considered here; ii) nuclei in this range can
produce substantial neutrino fluxes; iii) Icecube and its DeepCore sub-array
neutrino detectors are sensitive to the details of the neutrino spectrum in the
energy range $10-1000$ GeV.
In method B, the charged pion spectrum is approximated by a radial scaling
\cite{Hillas79ICRC}, based on the apparent Feynman scaling violation at
$x_{R}\equiv{E}/(\sqrt{s}/2)\ll1$. The meson decay kinematics are well
established, and in method B we follow the formulation of \cite{Marscher+80}
\footnote{Noting a typo in this reference, where in his Eq.[14], second line,
$\eta_{\nu}$ should be replaced by $\xi$.}.

\begin{figure}  \includegraphics[width=0.8\columnwidth]{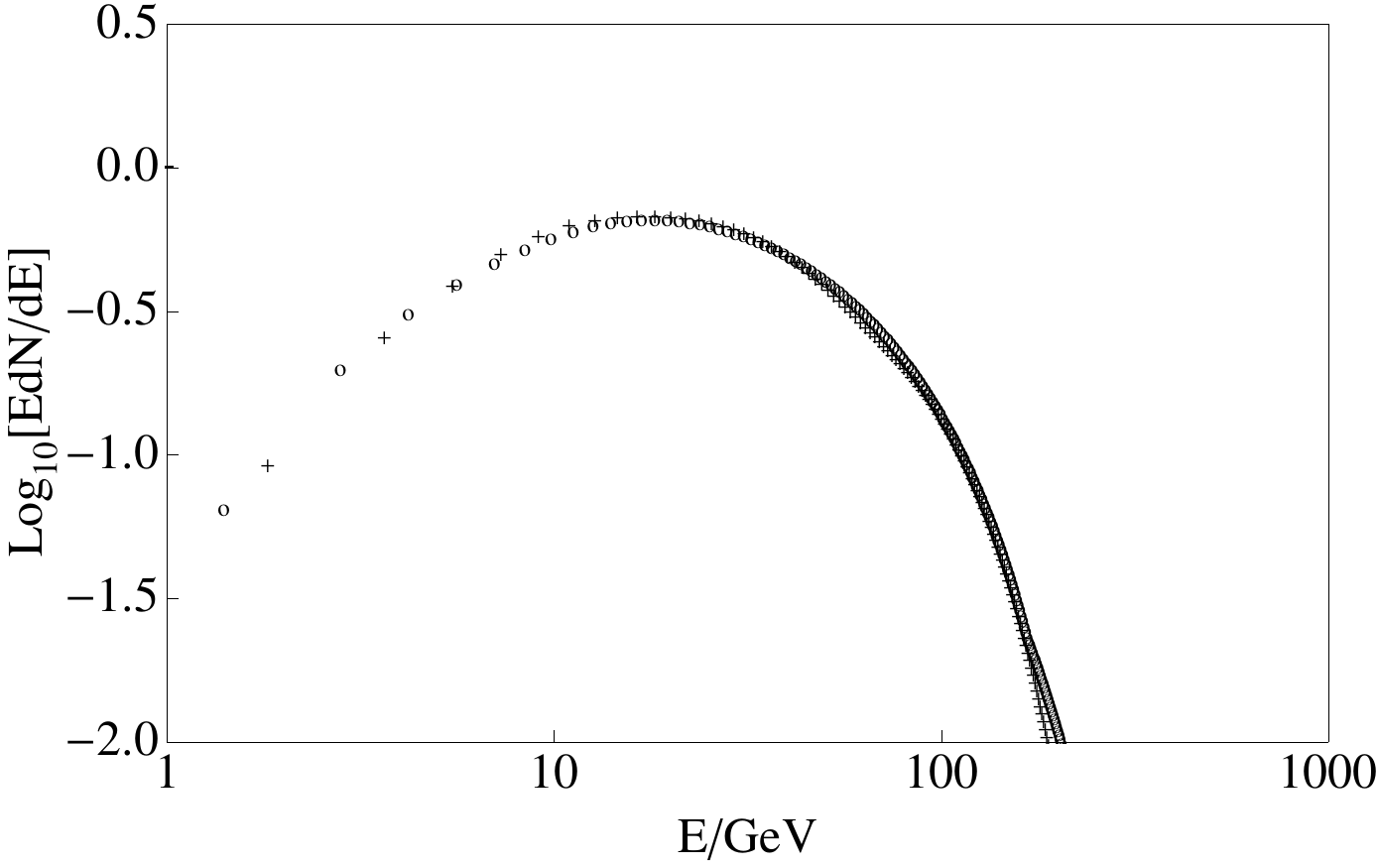}
\caption{The sum of the muon neutrino and anti-neutrino spectrum from nuclear
collisions, calculated in the stellar frame using the numerical method A (curve
marked by "+") and the method B (curve marked by "o"), as discussed in \S
\ref{sec:nusingle}. Both spectra are normalized to one $pp$ collision event.
In the outer jet comoving frame (where target protons are at rest) the incident
proton kinetic energy is $T_{p}=3.8$ GeV.  In the case shown here the outer jet
Lorentz factor is $\eta_{out}=70$ and the inner jet's is $\eta_{core}=700$.}
\label{AB} \end{figure}

A comparison of the muon neutrino and anti-neutrino spectra at the source
calculated using method A and B is plotted in Fig.[\ref{AB}], for an incident
proton energy $T_{p}=3.8$ GeV. It is seen that the two results agree  well with
each other. We note that a) both methods would result in a very small number of
of neutrinos which violate energy-momentum conservation. Although the total
amount of energy involved in these neutrinos is well below a fraction
$\sim10^{-3}$, and thus negligible, we have nonetheless applied a cutoff in the
spectrum beyond the energy where this occurs for both method A and B.  b) both
these methods simulate $pp$ collisions, instead of $pn$ collisions. For $pp$
collisions, the $\pi^{+}$ multiplicity is therefore greater than that of
$\pi^{-}$ due to charge conservation near the pion creation threshold energy.
At $\sqrt{s}\sim$ few GeV or higher, the two multiplicities tend to equal each
other. In method A we sum the neutrino and anti-neutrino from both $\pi^{+}$
and  $\pi^{-}$ channels.  In method B, Hillas used one fitted formula to
describe both $\pi^{+}$  and  $\pi^{-}$ spectra. Thus, for the summed neutrino
and anti-neutrino spectrum resulting from the decay products of these mesons,
the discrepancy between $pp$ and $pn$ collisions becomes less noticeable.  

\begin{figure}   \includegraphics[width=0.9\columnwidth]{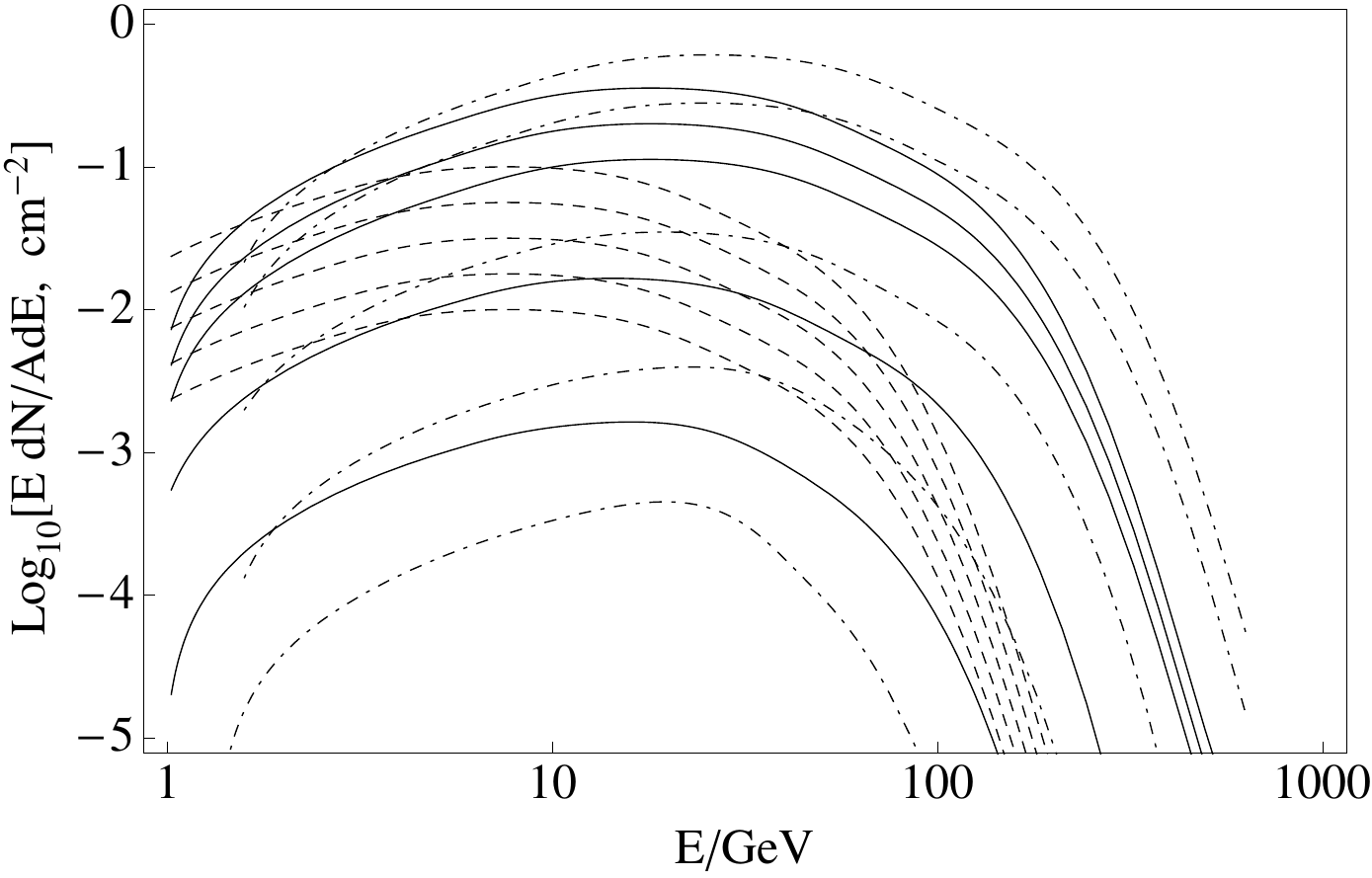} \caption{The
$\nu_{\mu}+{\bar{\nu}}_{\mu}$ fluence in the Earth observer frame from a single
burst, integrated over the outburst duration, for a nominal redshift of
$z=0.1$.  The other nominal parameters are the same as in Fig.[\ref{pionLT}],
except for $\eta_{core}= 10\eta_{out}= 300,700,1000$, indicated with dashed,
solid and dot-dashed lines respectively. The lines in each style are arranged
from top to bottom in the sequence $L_{tot}\sim L_{b}=10^{55},
10^{54.5},10^{54},10^{53.5},10^{53}$ erg/s.} \label{L}  \end{figure}

\begin{figure}   
\includegraphics[width=0.7\columnwidth]{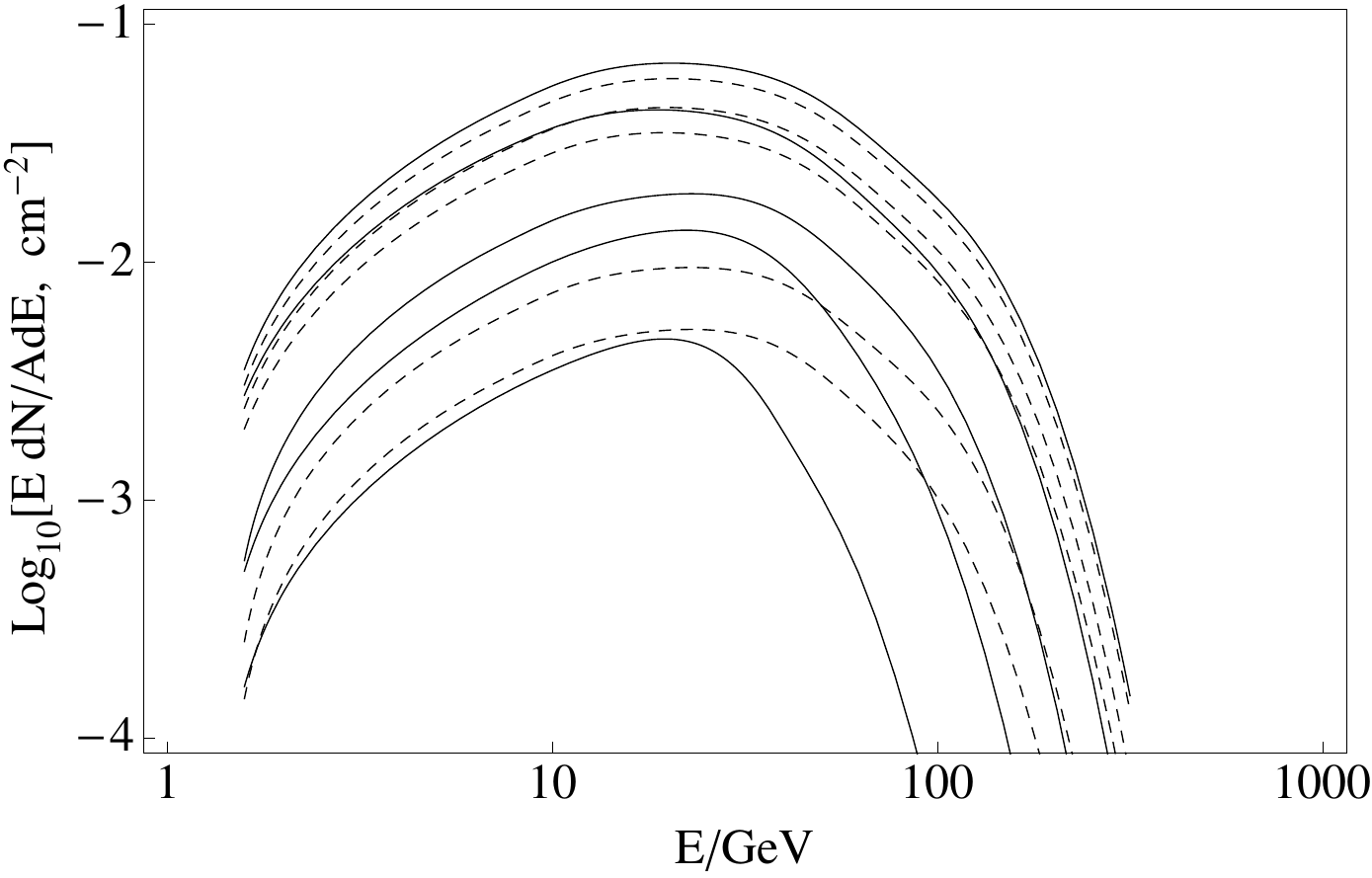}
\includegraphics[width=0.7\columnwidth]{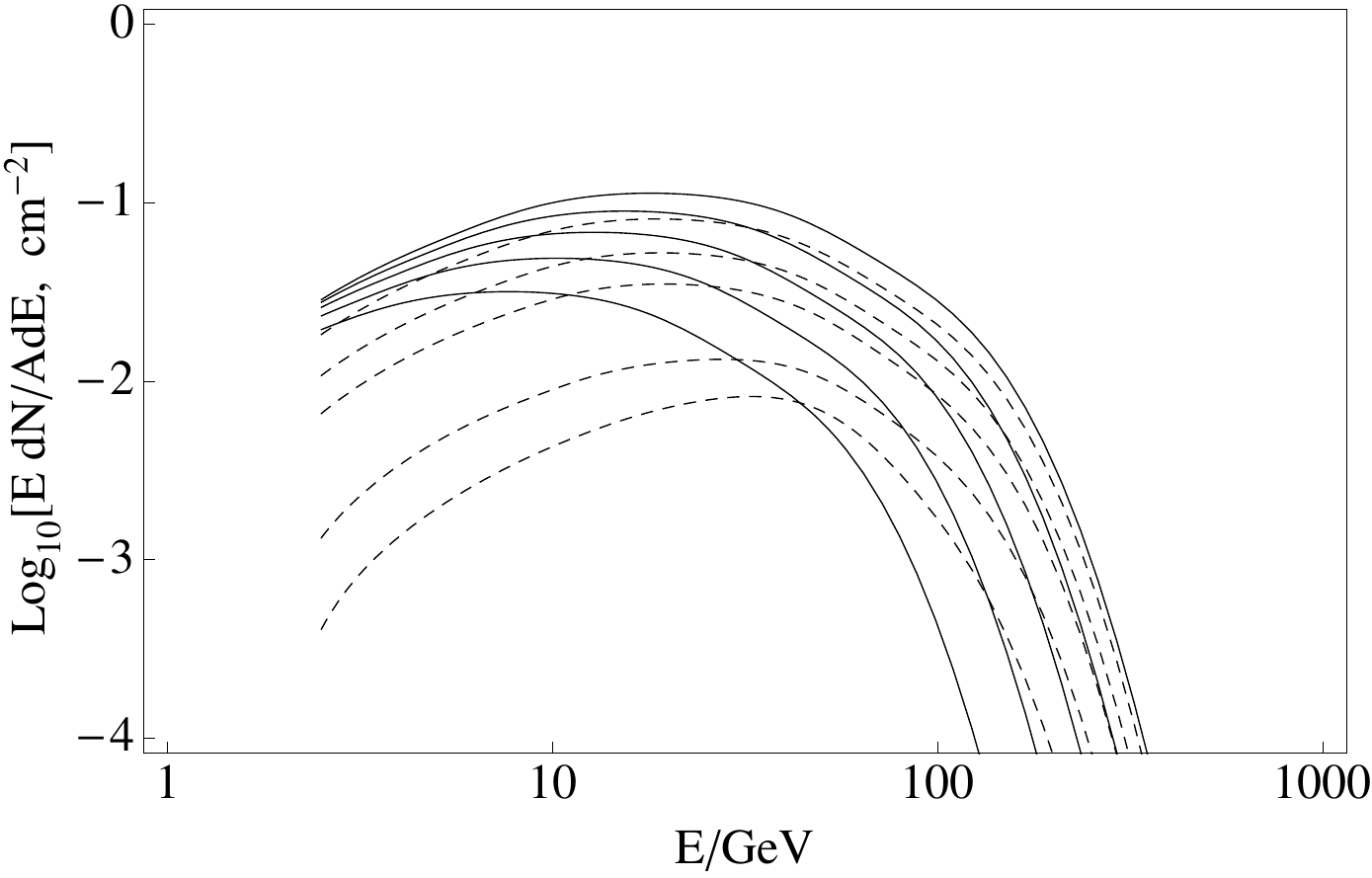} 
\caption{The
$\nu_{\mu}+{\bar{\nu}}_{\mu}$ fluence in the Earth observer frame from $z=0.1$,
for different parameters.  Top panel: solid lines from bottom to top are for
$\eta_{core}=300,400,500,600,700$; dashed lines from bottom to top are for
$\eta_{core}=2000,1500,1000,900,800$, with a fixed $\eta_{out}=100$ .  Bottom
panel: solid lines,  bottom to top are  $\eta_{core}=300,400,500,600,700$;
dashed lines, bottom to top are $\eta_{core}=1500,1200,1000,900,800$ with a
running $\eta_{out}=0.1\eta_{core}$.  Both panels are for $L_{p}=10^{54}$
erg/s, the value of $\eta_{core}$ yielding the largest neutrino flux being
around $\eta \sim 700$. When $\eta_{core}$ approaches 300 the pion multiplicity
becomes smaller due to a smaller relative Lorentz factor between outer and
inner jet particles (top figure, Eq. [\ref{eq:lambdapi}]) or due to a slightly
smaller diffusion rate (Eq. [\ref{eq:NN}]).  For very high $\eta_{core} \gtrsim
10^{3}$, the pion multiplicity similarly decreases.} \label{eta}  
\end{figure}

\begin{figure}   \includegraphics[width=0.9\columnwidth]{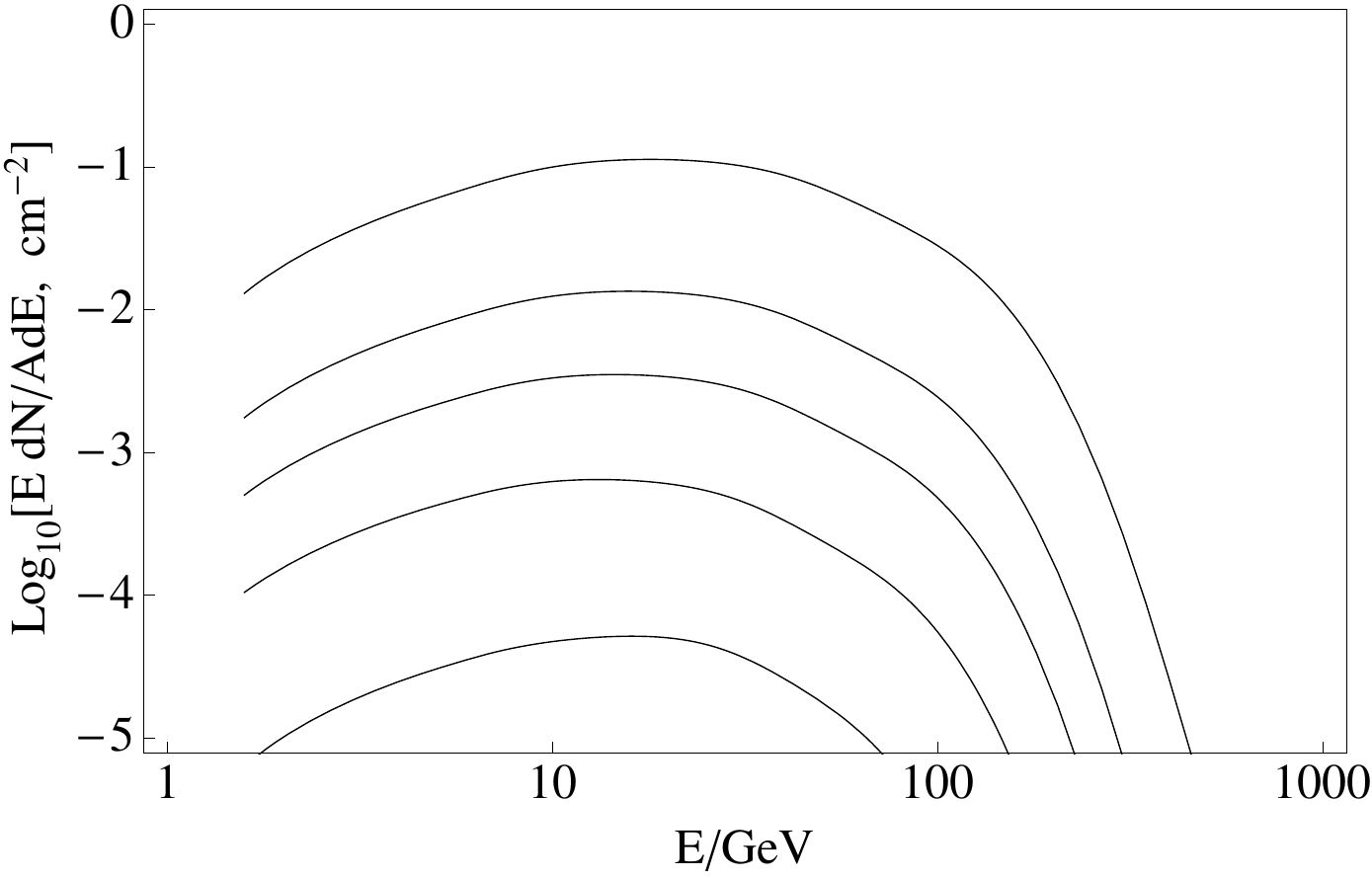}
\caption{The $\nu_{\mu}+{\bar{\nu}}_{\mu}$ number fluence in the Earth observer
frame from $z=0.1$, calculated for $\eta_{core}=10\eta_{out}=700$, $L_{T}\sim
L_{b}= 10^{54}erg/s$ for various jet opening angles
$\theta_{jet}=0.01,0.02,0.03, 0.04,0.05,0.1$ (top to bottom). } \label{theta}
\end{figure}

The muon neutrino and anti-neutrino differential number fluence at Earth for a
single burst (neutrinos per energy decade, integrated over the duration of the
outburst) is shown, after oscillations, in Figs.
[\ref{L},\ref{eta},\ref{theta}].  In Figure [\ref{L},\ref{eta}], we note a 
saturation effect in the dependence of $L_{\nu}$ on $L_{p}$ : at low $L_{p}$, 
$L_{\nu}$ grows fast with it and then stabilizes above $L_{p}\sim10^{54}$ erg/s.
This effect is mainly due to the change of relative Lorentz factors between
$\eta_{core}$ and $\eta_{out}$ by choices of different astrophysical parameters.
$\Gamma_{rel}$ grows with $L_p$ if we fix other parameters, and saturates at
some $L_{p,0}$ level (may be already saturated at the low end of $L_p$ in parts
of our parameter space.  A higher $\Gamma_{rel}$ is associated with more
secondary leptons (including neutrinos) per pn collision.  However when
saturated, the received neutrino flux grows at a slower rate with $L_p$,
because the increase of fluence is then only due to the fact that we have more
protons in the jet.

As discussed in \S 2, most of these neutrinos
come from transverse drift collisions, rather than from radial drifts. We have
used parameters representative of standard long GRB, extending from moderate to
high intrinsic luminosity, and the fluence is calculated for a nominal redshift
of $z=0.1$, corresponding to a luminosity distance of 450 Mpc in a standard
$\Omega_m=0.28$, $\Omega_V=0.72$, $h=0.72$ cosmology. This is at the lower end
of the classical redshift distribution, since for the 5-100 GeV neutrino
energies considered here only rare, very nearby bursts might be expected to be
detectable individually, due to the detector effective area decrease with
energy.  The typical neutrino fluences in Figs.  [\ref{L},\ref{eta},\ref{theta}]
are $\lesssim$ GeV cm$^{-2}\sim1.6\times10^{-3}$ erg cm$^{-2}$ for 
$z=0.1$. Since $pp$ collisions are expected to produce a comparable amount of 
energy in photons from $\pi^0$ decay as in neutrinos, it is important to check 
that such a photon luminosity
\footnote{An evaluation of the final photon spectrum would require the use of a detailed
electromagnetic cascade code, which is beyond the scope of this paper.}  
does not violate 
electromagnetic observation constraints. There are no gamma-ray detections at 
TeV energies so far, and the smattering of GeV detections involve bursts typically at 
much higher redshifts than considered here, so the possible constraints are 
mainly the 20-300 keV  energy range fluence statistics from the BATSE 4B
\cite{Paciesas+99} compilation of bursts. We can conservatively assume that at most 
a fraction $\lesssim 0.3$ of the total $\pi^0$ photon energy will appear in the 20-300 
keV range, i.e. $\lesssim5\times10^{-4}$ erg cm$^{-2}$. Such fluences are marginally
compatible with the BATSE 4B statistics, which does include some objects with fluences 
$\lesssim 10^{-3}$ erg/cm$^{2}$ (BATSE does not provide redshift information, but it is 
known that most BATSE bursts are at $z\gtrsim 1$; here we considered our bursts at $z<0.1$, 
and if these were placed at the typical $z>1$, they would show a BATSE fluence 
$\lesssim 5\times 10^{-6}$ erg cm$^{-2}$). Thus, our electromagnetic fluences appear 
compatible with the current observations.

The number of muon events arising from the above incident muon neutrino and
anti-neutrino fluences is calculated based on the specific detector
characteristics.  For the Icecube full 86-string operation, the effective area
is larger than that of Deep Core at $\sim100$ GeV, these effective areas being
given by \cite{IcecubeDC+09}. For a detection relying exclusively on neutrinos
one would need to consider GRBs which show at least one muon event. However,
bursts able to give an average of $>0.01$ muon events are also of interest,
because of the expectation of natural fluctuations in the distance or in the
luminosity, and because in some cases temporally coincident electromagnetic
observations can be expected.  The burst values of $L_\gamma$ and redshift $z$
yielding different muon event numbers for different burst parameters are shown
by the contours in Fig.[\ref{contour}].

The average rate per year at which GRBs occur producing $\geq 1$ muon events
can be estimated using the GRB luminosity and redshift distributions discussed
in \S \ref{sec:nudiff}.  For a baryon to photon luminosity ratio of 10, i.e.
$\epsilon_{b}=10\epsilon_{e} \approx1$, with inner to outer jet Lorentz factor
contrasts of $\eta_{core}= 10\eta_{out} = 300,700,1000$, respectively, this
average rate is expected to be around $0.0006/yr$, $0.04/yr$ and $0.06/yr$.

\subsection{Diffuse Neutrino Flux} \label{sec:nudiff}

The diffuse neutrino flux from all GRBs in the sky can be calculated using a
GRB luminosity distribution (luminosity function, LF) and a redshift
distribution (RD)\footnote{For the method of calculation, see e.g. Appendix B
of \cite{Murase08revisited}}. Here we adopt for these the functions given in
\cite{Wanderman+10grbsfr}, 
\begin{equation} \phi(L_{\gamma})\propto\begin{cases} (L_{\gamma}/L_{*})^{m_{1}}
& L_{min}<L_{\gamma}<L_{*}\\ (L_{\gamma}/L_{*})^{m_{2}} &
L_{*}<L_{\gamma}<L_{max} \end{cases} \label{eq:LF} \end{equation}
\begin{equation} R_{GRB}(z)\propto\begin{cases} (1+z)^{n_{1}} & z<z_{1}\\
(1+z)^{n_{2}} & z>z_{1} \end{cases}~.  \label{eq:RD} \end{equation}
Here equation [\ref{eq:LF}] is the luminosity function, and Eqn. [\ref{eq:RD}] is
the redshift distribution function, $L_{\gamma}$ is the peak photon luminosity
(typically in the 0.1-0.3 MeV range), $L_{min}=10^{50}$ erg/s,
$L_{*}=10^{52.5}$ erg/s, $L_{max}= 10^{54}$ erg/s, 
$m_{1}=-0.2$ , $m_{2}=-1.4$ , $n_{1}=1.0$ , $n_{2}=-1.4$ ,  
$z_{1}=3$ , and we have used
these values from the \cite{Wanderman+10grbsfr} parameter ranges which best
reproduce the actual GRBs with measured $L_\gamma$ and $z$ statistics, e.g.
Fig.2 of \cite{Wanderman+10grbsfr}, Fig.4 of \cite{Gehrels09rev}, or
\cite{GreinerWEBPAGE}.  We note that, especially in the range $z\lesssim 0.5$,
the index of the redshift distribution, i.e. the rate, is very uncertain, due
to poor statistics.  The differential co-moving rate of GRBs at a redshift $z$
is given by 
\begin{equation} R(z)=\frac{R_{GRB}(z)}{(1+z)}\frac{dV(z)}{dz} \end{equation}
where $V(z)$ is the comoving volume in the $\Lambda CDM$ cosmology model
adopted, with $\Omega_m=0.28$, $\Omega_V=0.72$ and $H_0=72$ Km/s/Mpc.  The
differential number of GRB per unit redshifts is given by 
\begin{equation} dN=\rho_{0}\phi(L_{\gamma})R(z)dlogL_{\gamma}dz \end{equation}
where $\rho_{0}$ is the normalization factor. In this paper, we normalize the
total electromagnetically detected GRB rate to $300/yr$ in the range given by
Eqns.  [\ref{eq:LF},\ref{eq:RD}].  The neutrino flux depends on the baryon
luminosity $L_{b}$, and the ratio between $L_{b}$ and $L_{\gamma}$ adopted in
\S \ref{sec:nusingle} is $L_{b}/L_{\gamma}\approx10$.  This ratio is
characteristic of many hadronic GRB models,  the implied photon radiative
efficiency of 10\% being moderate. The resulting diffuse neutrino fluxes
(neutrinos per year) are shown in Fig.[\ref{diff}], and are discussed in  \S
\ref{sec:disc}.

\section{Discussion}
\label{sec:disc}

We have calculated the neutrino emission in the range of a few GeV to a few
hundred GeV arising in magnetized collisional GRB models, such as have been
recently used for interpreting the electromagnetic properties of these objects.
The neutrino emission considered here arises partly from longitudinal proton
and neutron collisions following `their decoupling in the jet, and in larger
part from collisions caused by transverse thermal drift of neutrons from an
outer jet sheath into the jet core, having different bulk Lorentz factors. This
neutrino production model differs from the commonly considered photohadronic
$p\gamma$ models, and it also differs from previous $pn$ model calculations in
incorporating explicitly the (dominant) transverse drift collision effects.
Also, the emission region characteristics are here determined by the
magnetically dominated dynamics, differing from those in previous neutrino
calculations, which mostly use baryonic dynamics.  Furthermore, the $pn$
neutrino spectra are calculated numerically, using two different codes which
are suited for the GeV range energies considered.

Our present results indicate that for the burst parameters suitable to explain
the photon spectra and the MeV-GeV photon lags \cite{Meszaros+11gevmag}
indicated by the recent {\it Fermi} satellite observations, a low level of
neutrino emission is expected at $\sim 5-100$ GeV energies. This is in the
sensitivity range of the Deep Core sub-array of IceCube, and extends into the
lower range of the main IceCube array.

For a neutrino detection of an individual burst, unaided by a coincident
electromagnetic detection, one would require $\gtrsim 1$ muon event. For
neutrino-induced muon track events in IceCube and Deep Core, considering the
angular resolution
(e.g. $\Omega \lesssim 10^2~{\rm deg}^{2}$)
and time search bins $\sim o(1)\times20s$ per burst, the atmospheric neutrino 
background is low.  During the total sum of the search bins (e.g. assuming 
300 electromagnetically observed long GRBs in a year, 
from the atmospheric neutrino spectrum the expected number of
background muon events is $\lesssim o(1)\times0.09$.

\begin{figure}   
\includegraphics[width=0.95\columnwidth]{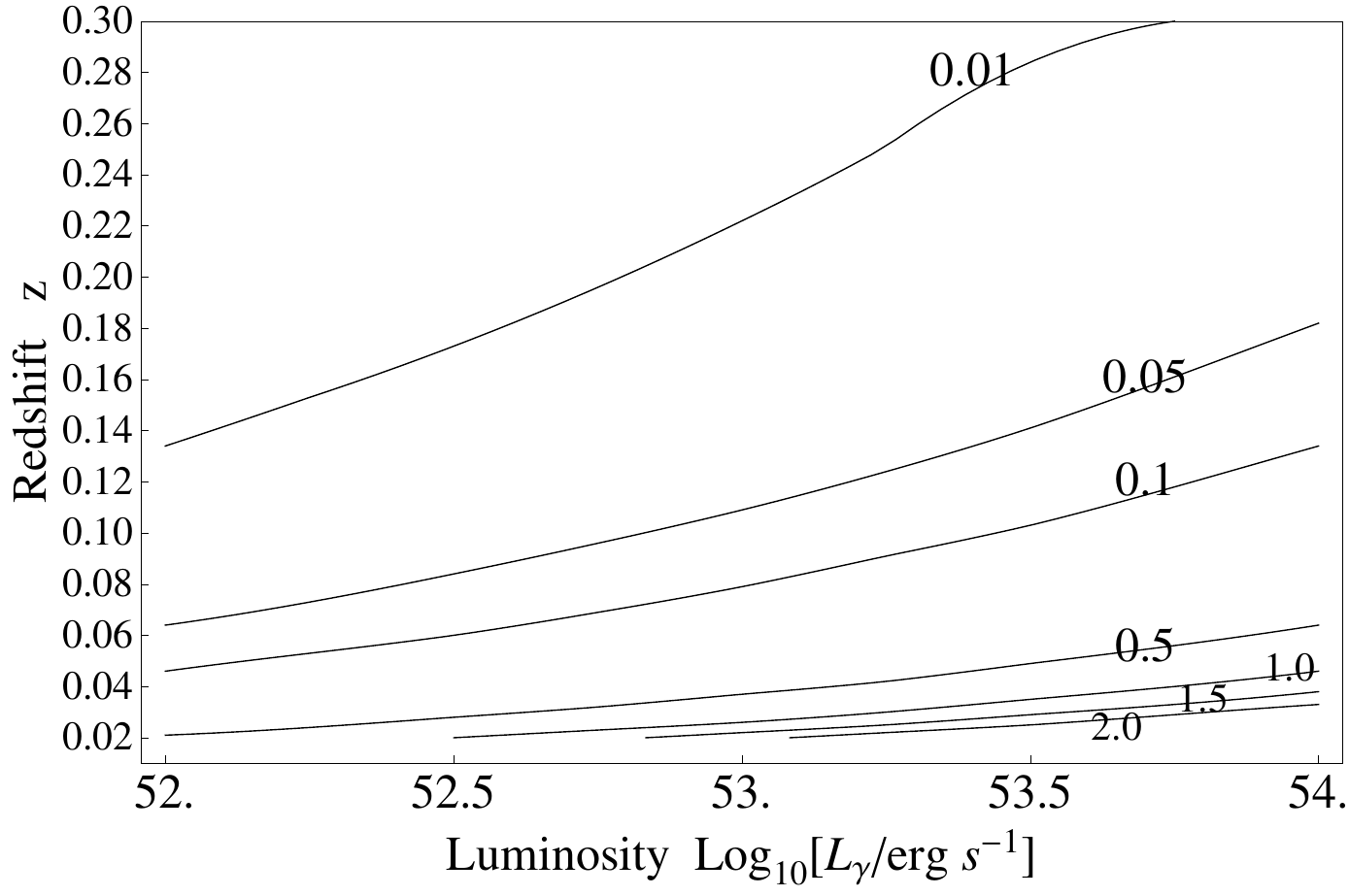}
\includegraphics[width=0.95\columnwidth]{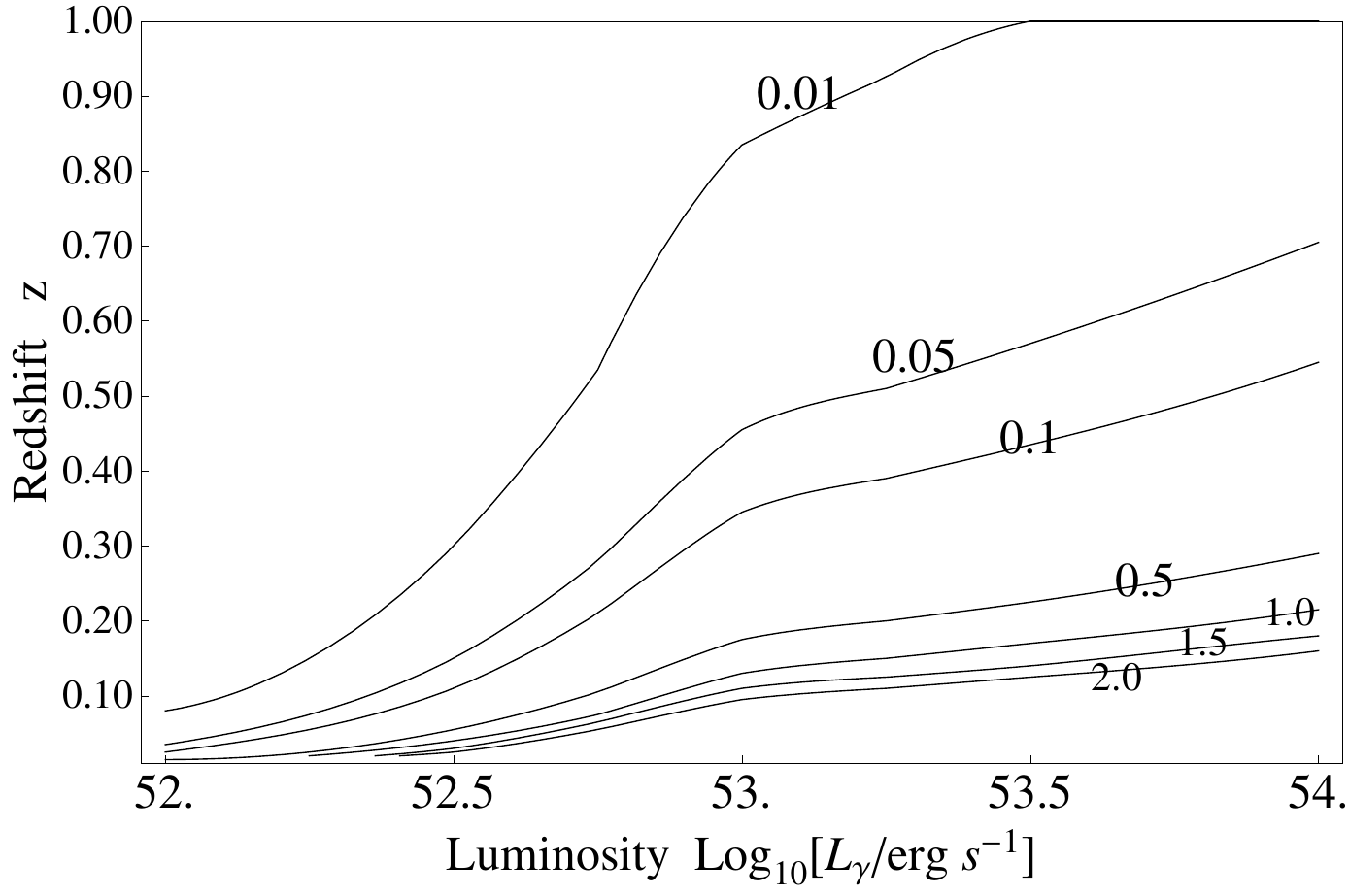}
\includegraphics[width=0.95\columnwidth]{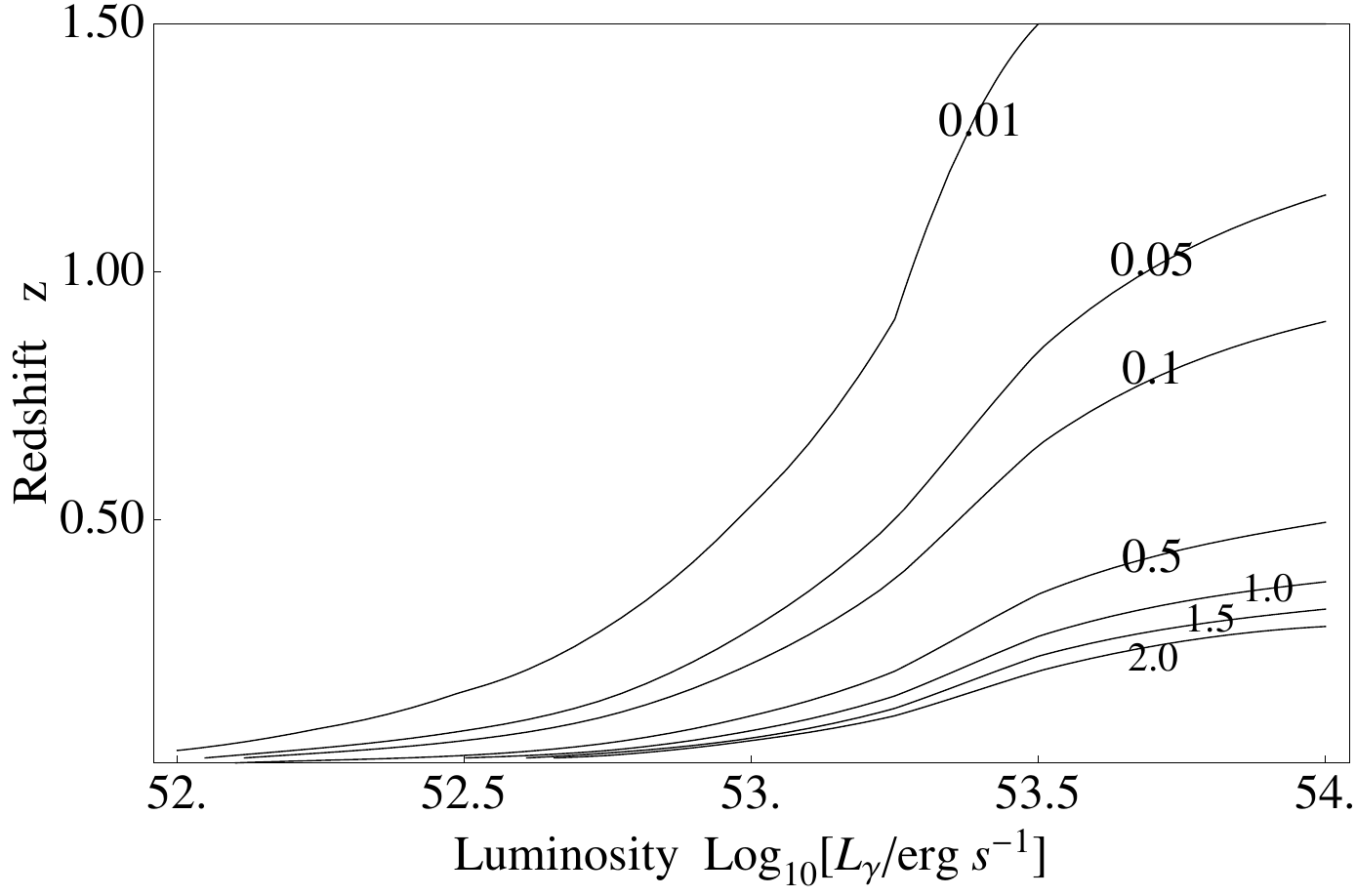}

\caption{Expected number of
muon events in the full 86-strings IceCube and its Deep Core sub-array, from a
\emph{single GRB} with $\eta_{core}=10\eta_{out}=300$(top panel),$700$(middle panel),
$1000$(bottom panel), 
from various redshifts
$z$ and for different photon luminosities $L_{\gamma}$, assuming a baryon to
photon luminosity ratio of 10, or $\epsilon_{b}=2\epsilon_{p}\sim10
\epsilon_{\gamma}$.  The effective detection areas are taken from
\cite{IcecubeDC+09}, with the angular position averaged over the northern sky
(the effective areas have some dependence on the incident angle of the
neutrinos). These contours show the $L_{\gamma}$ and $z$ ranges that give
$0.01-2$ muon events} \label{contour}   \end{figure}

However, any GRBs which might be expected to yield $\gtrsim1$ muon event need
to be at the high end of the luminosity function, and located at very low
redshifts, $L_{\gamma,iso} \gtrsim 10^{53}$ erg/s and $z\lesssim 0.1$.  From
the discussion of \S \ref{sec:nudiff}, the occurrence of such GRB is estimated
to be very rare, $\lesssim 1/17$ per year.  This estimate is uncertain, because
of the poor statistics in the determination of the redshift distribution in
this range.  Such bursts would in general also be detectable by photon
detectors such as {\it Swift} and {\it Fermi}, except for Earth occultations or
possible outages.  For the more frequent weaker or more distant bursts, taking
into account fluctuations in the average quantities, a neutrino observation
correlated with a photon detection can narrow the time bin search, increasing
the effective sensitivity of the detection.  The rate of occurrence of such
lower fluence GRBs can be calculated from the luminosity function and redshift
distribution, and is shown in Fig.  [\ref{contour}]).

\begin{figure}  
\includegraphics[width=0.8\columnwidth]{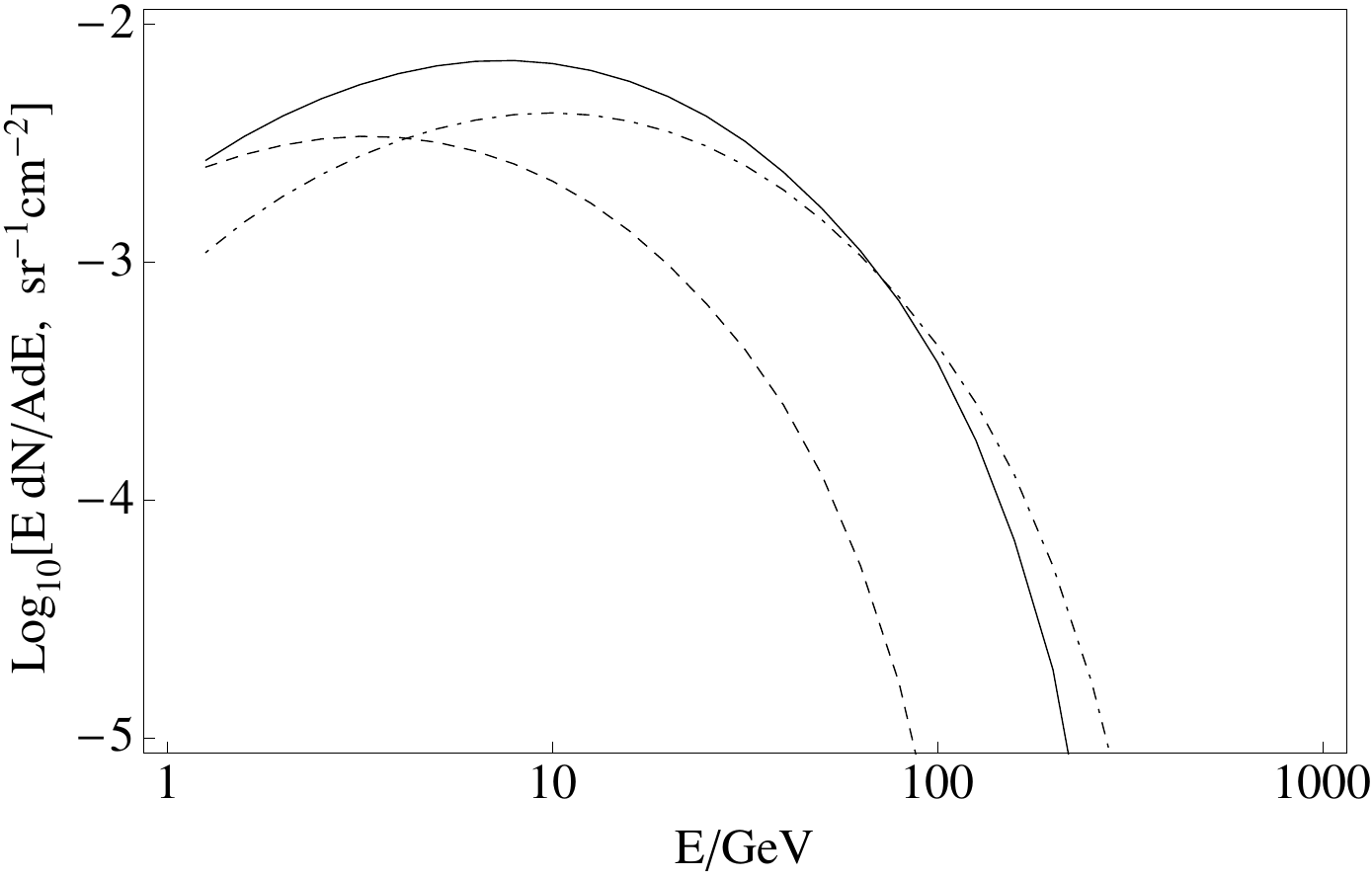}
\caption{The diffuse muon neutrino and anti-neutrino fluence per year (after
oscillations) for all GRBs down to $z>0.01$, as discussed in \S \ref{sec:disc},
using the luminosity function and redshift distribution of \S \ref{sec:nudiff}.
The conventions and parameters are the same as in Fig.[\ref{contour}], except for
using $\eta_{core}=10\eta_{out} =300$ (dashed), $700$ (solid) and $1000$
(dot-dashed). } \label{diff}   \end{figure}

Because of the low occurrence rate of bursts which can be expected to be
detected individually, it is useful to consider also the diffuse neutrino
fluxes. The same luminosity function and redshift distribution as above were
used for this, as discussed in \S \ref{sec:nudiff}, extending the integration
to all bursts with $z\ge 0.01$.  In Fig.[\ref{diff}] we show the diffuse
neutrino fluences over the period of a year, for a set of burst parameters
$\eta_{core}=10\eta_{out}=300,700,1000$ and
$\epsilon_{b}=10\epsilon_{e}\approx1$. For these paramaters, the corresponding
number of average expected muon events in Icecube and its DeepCore array are
estimated to be $0.03$, $0.4$ and $0.5$ per year over the whole sky,
respectively. Due to the poor statistics in the rate of low redshift
GRBs, the uncertainty in these numbers could be a factor $\sim 2$. The Icecube
IC40+IC59 muon event rates per year being used to set upper limits on the
TeV-PeV neutrino flux from putative Waxman-Bahcall GRB models (different from
the present model) are larger than the the event rates discussed here, but
within one order of magnitude of those for the highest $\eta$ values.  Thus,
imposing weak limits may perhaps be possible in the long term.

\medskip

In conclusion, both the individual burst fluences and the expected diffuse
flux in the 5-100 GeV range are significantly low. Fortuitous fluctuations
above the mean values could increase this somewhat, but any conclusions based
on the current IceCube and Deep Core arrays are likely to require years of data
accumulation with the full array. The proposed upgrades to these installations
would help, but a next generation of larger effective volume neutrino detectors
could be the best way to accelerate the detection or non-detection of GRB
neutrinos in this energy range, and to test GRB models such as discussed here.

\medskip

We thank P. Veres, M. Strikman, M. Smith, D. Cowen and especially T. DeYoung
for useful discussions, the referee for valuable critical comments, and NASA
NNX09AL40G, NSF PHY-0757155 for partial support.

\newpage
\bibliography{grbMG}

\begin{thebibliography}{39}%
\makeatletter
\providecommand \@ifxundefined [1]{%
 \@ifx{#1\undefined}
}%
\providecommand \@ifnum [1]{%
 \ifnum #1\expandafter \@firstoftwo
 \else \expandafter \@secondoftwo
 \fi
}%
\providecommand \@ifx [1]{%
 \ifx #1\expandafter \@firstoftwo
 \else \expandafter \@secondoftwo
 \fi
}%
\providecommand \natexlab [1]{#1}%
\providecommand \enquote  [1]{``#1''}%
\providecommand \bibnamefont  [1]{#1}%
\providecommand \bibfnamefont [1]{#1}%
\providecommand \citenamefont [1]{#1}%
\providecommand \href@noop [0]{\@secondoftwo}%
\providecommand \href [0]{\begingroup \@sanitize@url \@href}%
\providecommand \@href[1]{\@@startlink{#1}\@@href}%
\providecommand \@@href[1]{\endgroup#1\@@endlink}%
\providecommand \@sanitize@url [0]{\catcode `\\12\catcode `\$12\catcode
  `\&12\catcode `\#12\catcode `\^12\catcode `\_12\catcode `\%12\relax}%
\providecommand \@@startlink[1]{}%
\providecommand \@@endlink[0]{}%
\providecommand \url  [0]{\begingroup\@sanitize@url \@url }%
\providecommand \@url [1]{\endgroup\@href {#1}{\urlprefix }}%
\providecommand \urlprefix  [0]{URL }%
\providecommand \Eprint [0]{\href }%
\providecommand \doibase [0]{http://dx.doi.org/}%
\providecommand \selectlanguage [0]{\@gobble}%
\providecommand \bibinfo  [0]{\@secondoftwo}%
\providecommand \bibfield  [0]{\@secondoftwo}%
\providecommand \translation [1]{[#1]}%
\providecommand \BibitemOpen [0]{}%
\providecommand \bibitemStop [0]{}%
\providecommand \bibitemNoStop [0]{.\EOS\space}%
\providecommand \EOS [0]{\spacefactor3000\relax}%
\providecommand \BibitemShut  [1]{\csname bibitem#1\endcsname}%
\let\auto@bib@innerbib\@empty
\bibitem [{\citenamefont {{Abdo}}\ and\ \citenamefont {the {Fermi
  collab.}}(2009)}]{Abdo+09grb080916c}%
  \BibitemOpen
  \bibfield  {author} {\bibinfo {author} {\bibfnamefont {A.~A.}\ \bibnamefont
  {{Abdo}}}\ and\ \bibinfo {author} {\bibnamefont {the {Fermi collab.}}},\
  }\href {\doibase 10.1126/science.1169101} {\bibfield  {journal} {\bibinfo
  {journal} {Science}\ }\textbf {\bibinfo {volume} {323}},\ \bibinfo {pages}
  {1688} (\bibinfo {year} {2009})}\BibitemShut {NoStop}%
\bibitem [{\citenamefont {{Ackermann}}\ and\ \citenamefont {the
  Fermi~collab.}(2010)}]{Ackermann+10grb090510}%
  \BibitemOpen
  \bibfield  {author} {\bibinfo {author} {\bibfnamefont {M.}~\bibnamefont
  {{Ackermann}}}\ and\ \bibinfo {author} {\bibnamefont {the Fermi~collab.}},\
  }\href {\doibase 10.1088/0004-637X/716/2/1178} {\bibfield  {journal}
  {\bibinfo  {journal} {\apj}\ }\textbf {\bibinfo {volume} {716}},\ \bibinfo
  {pages} {1178} (\bibinfo {year} {2010})}\BibitemShut {NoStop}%
\bibitem [{\citenamefont {{Pe'er}}(2011)}]{Peer11-fermigrb}%
  \BibitemOpen
  \bibfield  {author} {\bibinfo {author} {\bibfnamefont {A.}~\bibnamefont
  {{Pe'er}}},\ }\href@noop {} {\bibfield  {journal} {\bibinfo  {journal} {ArXiv
  e-prints}\ } (\bibinfo {year} {2011})},\ \Eprint
  {http://arxiv.org/abs/1111.3378} {arXiv:1111.3378 [astro-ph.HE]} \BibitemShut
  {NoStop}%
\bibitem [{\citenamefont {{M{\'e}sz{\'a}ros}}\ and\ \citenamefont
  {{Rees}}(2000{\natexlab{a}})}]{Meszaros+00phot}%
  \BibitemOpen
  \bibfield  {author} {\bibinfo {author} {\bibfnamefont {P.}~\bibnamefont
  {{M{\'e}sz{\'a}ros}}}\ and\ \bibinfo {author} {\bibfnamefont
  {M.}~\bibnamefont {{Rees}}},\ }\href {\doibase 10.1086/308371} {\bibfield
  {journal} {\bibinfo  {journal} {\apj}\ }\textbf {\bibinfo {volume} {530}},\
  \bibinfo {pages} {292} (\bibinfo {year} {2000}{\natexlab{a}})},\ \Eprint
  {http://arxiv.org/abs/arXiv:astro-ph/9908126} {arXiv:astro-ph/9908126}
  \BibitemShut {NoStop}%
\bibitem [{\citenamefont {{Rees}}\ and\ \citenamefont
  {{M{\'e}sz{\'a}ros}}(2005)}]{Rees+05photdis}%
  \BibitemOpen
  \bibfield  {author} {\bibinfo {author} {\bibfnamefont {M.~J.}\ \bibnamefont
  {{Rees}}}\ and\ \bibinfo {author} {\bibfnamefont {P.}~\bibnamefont
  {{M{\'e}sz{\'a}ros}}},\ }\href {\doibase 10.1086/430818} {\bibfield
  {journal} {\bibinfo  {journal} {\apj}\ }\textbf {\bibinfo {volume} {628}},\
  \bibinfo {pages} {847} (\bibinfo {year} {2005})},\ \Eprint
  {http://arxiv.org/abs/arXiv:astro-ph/0412702} {arXiv:astro-ph/0412702}
  \BibitemShut {NoStop}%
\bibitem [{\citenamefont {{Beloborodov}}(2009)}]{Beloborodov10phot}%
  \BibitemOpen
  \bibfield  {author} {\bibinfo {author} {\bibfnamefont {A.~M.}\ \bibnamefont
  {{Beloborodov}}},\ }\href@noop {} {\bibfield  {journal} {\bibinfo  {journal}
  {ArXiv e-prints}\ } (\bibinfo {year} {2009})},\ \Eprint
  {http://arxiv.org/abs/0907.0732} {arXiv:0907.0732 [astro-ph.HE]} \BibitemShut
  {NoStop}%
\bibitem [{\citenamefont {{Derishev}}\ \emph {et~al.}(1999)\citenamefont
  {{Derishev}}, \citenamefont {{Kocharovsky}},\ and\ \citenamefont
  {{Kocharovsky}}}]{Derishev+99}%
  \BibitemOpen
  \bibfield  {author} {\bibinfo {author} {\bibfnamefont {E.~V.}\ \bibnamefont
  {{Derishev}}}, \bibinfo {author} {\bibfnamefont {V.~V.}\ \bibnamefont
  {{Kocharovsky}}}, \ and\ \bibinfo {author} {\bibfnamefont {V.~V.}\
  \bibnamefont {{Kocharovsky}}},\ }\href {\doibase 10.1086/307574} {\bibfield
  {journal} {\bibinfo  {journal} {\apj}\ }\textbf {\bibinfo {volume} {521}},\
  \bibinfo {pages} {640} (\bibinfo {year} {1999})}\BibitemShut {NoStop}%
\bibitem [{\citenamefont {{Bahcall}}\ and\ \citenamefont
  {{M{\'e}sz{\'a}ros}}(2000)}]{Bahcall+00pn}%
  \BibitemOpen
  \bibfield  {author} {\bibinfo {author} {\bibfnamefont {J.~N.}\ \bibnamefont
  {{Bahcall}}}\ and\ \bibinfo {author} {\bibfnamefont {P.}~\bibnamefont
  {{M{\'e}sz{\'a}ros}}},\ }\href@noop {} {\bibfield  {journal} {\bibinfo
  {journal} {Physical Review Letters}\ }\textbf {\bibinfo {volume} {85}},\
  \bibinfo {pages} {1362} (\bibinfo {year} {2000})},\ \Eprint
  {http://arxiv.org/abs/arXiv:hep-ph/0004019} {arXiv:hep-ph/0004019}
  \BibitemShut {NoStop}%
\bibitem [{\citenamefont {{M{\'e}sz{\'a}ros}}\ and\ \citenamefont
  {{Rees}}(2000{\natexlab{b}})}]{Meszaros+00gevnu}%
  \BibitemOpen
  \bibfield  {author} {\bibinfo {author} {\bibfnamefont {P.}~\bibnamefont
  {{M{\'e}sz{\'a}ros}}}\ and\ \bibinfo {author} {\bibfnamefont {M.~J.}\
  \bibnamefont {{Rees}}},\ }\href {\doibase 10.1086/312894} {\bibfield
  {journal} {\bibinfo  {journal} {\apjl}\ }\textbf {\bibinfo {volume} {541}},\
  \bibinfo {pages} {L5} (\bibinfo {year} {2000}{\natexlab{b}})},\ \Eprint
  {http://arxiv.org/abs/arXiv:astro-ph/0007102} {arXiv:astro-ph/0007102}
  \BibitemShut {NoStop}%
\bibitem [{\citenamefont {{Giannios}}\ and\ \citenamefont
  {{Spruit}}(2007)}]{Giannios+07photspec}%
  \BibitemOpen
  \bibfield  {author} {\bibinfo {author} {\bibfnamefont {D.}~\bibnamefont
  {{Giannios}}}\ and\ \bibinfo {author} {\bibfnamefont {H.~C.}\ \bibnamefont
  {{Spruit}}},\ }\href {\doibase 10.1051/0004-6361:20066739} {\bibfield
  {journal} {\bibinfo  {journal} {\aap}\ }\textbf {\bibinfo {volume} {469}},\
  \bibinfo {pages} {1} (\bibinfo {year} {2007})},\ \Eprint
  {http://arxiv.org/abs/arXiv:astro-ph/0611385} {arXiv:astro-ph/0611385}
  \BibitemShut {NoStop}%
\bibitem [{\citenamefont {{Tchekhovskoy}}\ \emph {et~al.}(2010)\citenamefont
  {{Tchekhovskoy}}, \citenamefont {{Narayan}},\ and\ \citenamefont
  {{McKinney}}}]{Tchekhovskoy+10grb}%
  \BibitemOpen
  \bibfield  {author} {\bibinfo {author} {\bibfnamefont {A.}~\bibnamefont
  {{Tchekhovskoy}}}, \bibinfo {author} {\bibfnamefont {R.}~\bibnamefont
  {{Narayan}}}, \ and\ \bibinfo {author} {\bibfnamefont {J.~C.}\ \bibnamefont
  {{McKinney}}},\ }\href {\doibase 10.1016/j.newast.2010.03.001} {\ \textbf
  {\bibinfo {volume} {15}},\ \bibinfo {pages} {749} (\bibinfo {year} {2010})},\
  \Eprint {http://arxiv.org/abs/0909.0011} {arXiv:0909.0011 [astro-ph.HE]}
  \BibitemShut {NoStop}%
\bibitem [{\citenamefont {{M\'esz\'aros}}\ and\ \citenamefont
  {{Rees}}(2010)}]{Meszaros+10pop3}%
  \BibitemOpen
  \bibfield  {author} {\bibinfo {author} {\bibfnamefont {P.}~\bibnamefont
  {{M\'esz\'aros}}}\ and\ \bibinfo {author} {\bibfnamefont {M.}~\bibnamefont
  {{Rees}}},\ }\href {\doibase 10.1088/0004-637X/715/2/967} {\bibfield
  {journal} {\bibinfo  {journal} {\apj}\ }\textbf {\bibinfo {volume} {715}},\
  \bibinfo {pages} {967} (\bibinfo {year} {2010})},\ \Eprint
  {http://arxiv.org/abs/1004.2056} {arXiv:1004.2056 [astro-ph.HE]} \BibitemShut
  {NoStop}%
\bibitem [{\citenamefont {{Usov}}(1994)}]{Usov94}%
  \BibitemOpen
  \bibfield  {author} {\bibinfo {author} {\bibfnamefont {V.~V.}\ \bibnamefont
  {{Usov}}},\ }\href@noop {} {\bibfield  {journal} {\bibinfo  {journal}
  {\mnras}\ }\textbf {\bibinfo {volume} {267}},\ \bibinfo {pages} {1035}
  (\bibinfo {year} {1994})},\ \Eprint
  {http://arxiv.org/abs/arXiv:astro-ph/9312024} {arXiv:astro-ph/9312024}
  \BibitemShut {NoStop}%
\bibitem [{\citenamefont {{M\'esz\'aros}}\ and\ \citenamefont
  {{Rees}}(1997)}]{Meszaros+97poynting}%
  \BibitemOpen
  \bibfield  {author} {\bibinfo {author} {\bibfnamefont {P.}~\bibnamefont
  {{M\'esz\'aros}}}\ and\ \bibinfo {author} {\bibfnamefont {M.~J.}\
  \bibnamefont {{Rees}}},\ }\href@noop {} {\bibfield  {journal} {\bibinfo
  {journal} {\apjl}\ }\textbf {\bibinfo {volume} {482}},\ \bibinfo {pages}
  {L29+} (\bibinfo {year} {1997})},\ \Eprint
  {http://arxiv.org/abs/arXiv:astro-ph/9609065} {arXiv:astro-ph/9609065}
  \BibitemShut {NoStop}%
\bibitem [{\citenamefont {{Toma}}\ \emph {et~al.}(2011)\citenamefont {{Toma}},
  \citenamefont {{Sakamoto}},\ and\ \citenamefont
  {{M{\'e}sz{\'a}ros}}}]{Toma+11pop3}%
  \BibitemOpen
  \bibfield  {author} {\bibinfo {author} {\bibfnamefont {K.}~\bibnamefont
  {{Toma}}}, \bibinfo {author} {\bibfnamefont {T.}~\bibnamefont {{Sakamoto}}},
  \ and\ \bibinfo {author} {\bibfnamefont {P.}~\bibnamefont
  {{M{\'e}sz{\'a}ros}}},\ }\href {\doibase 10.1088/0004-637X/731/2/127}
  {\bibfield  {journal} {\bibinfo  {journal} {\apj}\ }\textbf {\bibinfo
  {volume} {731}},\ \bibinfo {pages} {127} (\bibinfo {year} {2011})},\ \Eprint
  {http://arxiv.org/abs/1008.1269} {arXiv:1008.1269 [astro-ph.CO]} \BibitemShut
  {NoStop}%
\bibitem [{\citenamefont {{Gao}}\ \emph {et~al.}(2011)\citenamefont {{Gao}},
  \citenamefont {{Toma}},\ and\ \citenamefont
  {{M{\'e}sz{\'a}ros}}}]{Gao+11pop3nu}%
  \BibitemOpen
  \bibfield  {author} {\bibinfo {author} {\bibfnamefont {S.}~\bibnamefont
  {{Gao}}}, \bibinfo {author} {\bibfnamefont {K.}~\bibnamefont {{Toma}}}, \
  and\ \bibinfo {author} {\bibfnamefont {P.}~\bibnamefont
  {{M{\'e}sz{\'a}ros}}},\ }\href {\doibase 10.1103/PhysRevD.83.103004}
  {\bibfield  {journal} {\bibinfo  {journal} {\prd}\ }\textbf {\bibinfo
  {volume} {83}},\ \bibinfo {pages} {103004} (\bibinfo {year} {2011})},\
  \Eprint {http://arxiv.org/abs/1103.5477} {arXiv:1103.5477 [astro-ph.HE]}
  \BibitemShut {NoStop}%
\bibitem [{\citenamefont {{Thompson}}(1994)}]{Thompson94}%
  \BibitemOpen
  \bibfield  {author} {\bibinfo {author} {\bibfnamefont {C.}~\bibnamefont
  {{Thompson}}},\ }\href@noop {} {\bibfield  {journal} {\bibinfo  {journal}
  {\mnras}\ }\textbf {\bibinfo {volume} {270}},\ \bibinfo {pages} {480}
  (\bibinfo {year} {1994})}\BibitemShut {NoStop}%
\bibitem [{\citenamefont {{Drenkhahn}}\ and\ \citenamefont
  {{Spruit}}(2002)}]{Drenkhahn+02}%
  \BibitemOpen
  \bibfield  {author} {\bibinfo {author} {\bibfnamefont {G.}~\bibnamefont
  {{Drenkhahn}}}\ and\ \bibinfo {author} {\bibfnamefont {H.~C.}\ \bibnamefont
  {{Spruit}}},\ }\href {\doibase 10.1051/0004-6361:20020839} {\bibfield
  {journal} {\bibinfo  {journal} {\aap}\ }\textbf {\bibinfo {volume} {391}},\
  \bibinfo {pages} {1141} (\bibinfo {year} {2002})},\ \Eprint
  {http://arxiv.org/abs/arXiv:astro-ph/0202387} {arXiv:astro-ph/0202387}
  \BibitemShut {NoStop}%
\bibitem [{\citenamefont {{Lyutikov}}\ and\ \citenamefont
  {{Blandford}}(2003)}]{Lyutikov+03grbmag}%
  \BibitemOpen
  \bibfield  {author} {\bibinfo {author} {\bibfnamefont {M.}~\bibnamefont
  {{Lyutikov}}}\ and\ \bibinfo {author} {\bibfnamefont {R.}~\bibnamefont
  {{Blandford}}},\ }\href@noop {} {\bibfield  {journal} {\bibinfo  {journal}
  {ArXiv Astrophysics e-prints}\ } (\bibinfo {year} {2003})},\ \Eprint
  {http://arxiv.org/abs/arXiv:astro-ph/0312347} {arXiv:astro-ph/0312347}
  \BibitemShut {NoStop}%
\bibitem [{\citenamefont {{McKinney}}\ and\ \citenamefont
  {{Uzdensky}}(2010)}]{McKinney+11magphot}%
  \BibitemOpen
  \bibfield  {author} {\bibinfo {author} {\bibfnamefont {J.~C.}\ \bibnamefont
  {{McKinney}}}\ and\ \bibinfo {author} {\bibfnamefont {D.~A.}\ \bibnamefont
  {{Uzdensky}}},\ }\href@noop {} {\bibfield  {journal} {\bibinfo  {journal}
  {ArXiv e-prints}\ } (\bibinfo {year} {2010})},\ \Eprint
  {http://arxiv.org/abs/1011.1904} {arXiv:1011.1904 [astro-ph.HE]} \BibitemShut
  {NoStop}%
\bibitem [{\citenamefont {{Metzger}}\ \emph
  {et~al.}(2011{\natexlab{a}})\citenamefont {{Metzger}}, \citenamefont
  {{Giannios}}, \citenamefont {{Thompson}}, \citenamefont {{Bucciantini}},\
  and\ \citenamefont {{Quataert}}}]{Metzger+11grbmag}%
  \BibitemOpen
  \bibfield  {author} {\bibinfo {author} {\bibfnamefont {B.~D.}\ \bibnamefont
  {{Metzger}}}, \bibinfo {author} {\bibfnamefont {D.}~\bibnamefont
  {{Giannios}}}, \bibinfo {author} {\bibfnamefont {T.~A.}\ \bibnamefont
  {{Thompson}}}, \bibinfo {author} {\bibfnamefont {N.}~\bibnamefont
  {{Bucciantini}}}, \ and\ \bibinfo {author} {\bibfnamefont {E.}~\bibnamefont
  {{Quataert}}},\ }\href {\doibase 10.1111/j.1365-2966.2011.18280.x} {\bibfield
   {journal} {\bibinfo  {journal} {\mnras}\ }\textbf {\bibinfo {volume}
  {413}},\ \bibinfo {pages} {2031} (\bibinfo {year} {2011}{\natexlab{a}})},\
  \Eprint {http://arxiv.org/abs/1012.0001} {arXiv:1012.0001 [astro-ph.HE]}
  \BibitemShut {NoStop}%
\bibitem [{\citenamefont {{Koers}}\ and\ \citenamefont
  {{Giannios}}(2007)}]{Giannios07pn}%
  \BibitemOpen
  \bibfield  {author} {\bibinfo {author} {\bibfnamefont {H.~B.~J.}\
  \bibnamefont {{Koers}}}\ and\ \bibinfo {author} {\bibfnamefont
  {D.}~\bibnamefont {{Giannios}}},\ }\href {\doibase
  10.1051/0004-6361:20077560} {\bibfield  {journal} {\bibinfo  {journal}
  {\aap}\ }\textbf {\bibinfo {volume} {471}},\ \bibinfo {pages} {395} (\bibinfo
  {year} {2007})},\ \Eprint {http://arxiv.org/abs/arXiv:astro-ph/0703719}
  {arXiv:astro-ph/0703719} \BibitemShut {NoStop}%
\bibitem [{\citenamefont {{M{\'e}sz{\'a}ros}}\ and\ \citenamefont
  {{Rees}}(2011)}]{Meszaros+11gevmag}%
  \BibitemOpen
  \bibfield  {author} {\bibinfo {author} {\bibfnamefont {P.}~\bibnamefont
  {{M{\'e}sz{\'a}ros}}}\ and\ \bibinfo {author} {\bibfnamefont
  {M.}~\bibnamefont {{Rees}}},\ }\href {\doibase 10.1088/2041-8205/733/2/L40}
  {\bibfield  {journal} {\bibinfo  {journal} {\apjl}\ }\textbf {\bibinfo
  {volume} {733}},\ \bibinfo {pages} {L40+} (\bibinfo {year} {2011})},\ \Eprint
  {http://arxiv.org/abs/1104.5025} {arXiv:1104.5025 [astro-ph.HE]} \BibitemShut
  {NoStop}%
\bibitem [{\citenamefont {{DeYoung}}\ and\ \citenamefont {{for the IceCube
  Collaboration}}(2011)}]{DeYoung+11RICAP}%
  \BibitemOpen
  \bibfield  {author} {\bibinfo {author} {\bibfnamefont {T.}~\bibnamefont
  {{DeYoung}}}\ and\ \bibinfo {author} {\bibnamefont {{for the IceCube
  Collaboration}}},\ }\href@noop {} {\bibfield  {journal} {\bibinfo  {journal}
  {ArXiv e-prints}\ } (\bibinfo {year} {2011})},\ \Eprint
  {http://arxiv.org/abs/1112.1053} {arXiv:1112.1053 [astro-ph.HE]} \BibitemShut
  {NoStop}%
\bibitem [{\citenamefont {{The IceCube Collaboration}.}(2011)}]{IcecubeDC+09}%
  \BibitemOpen
  \bibfield  {author} {\bibinfo {author} {\bibnamefont {{The IceCube
  Collaboration}.}},\ }\href@noop {} {\bibfield  {journal} {\bibinfo  {journal}
  {ArXiv e-prints}\ } (\bibinfo {year} {2011})},\ \Eprint
  {http://arxiv.org/abs/1109.6096} {arXiv:1109.6096 [astro-ph.IM]} \BibitemShut
  {NoStop}%
\bibitem [{\citenamefont {{The IceCube Collaboration}}(2011)}]{IcecubeICRC+11}%
  \BibitemOpen
  \bibfield  {author} {\bibinfo {author} {\bibnamefont {{The IceCube
  Collaboration}}},\ }\href@noop {} {\bibfield  {journal} {\bibinfo  {journal}
  {ArXiv e-prints}\ } (\bibinfo {year} {2011})},\ \Eprint
  {http://arxiv.org/abs/1111.5188} {arXiv:1111.5188 [astro-ph.HE]} \BibitemShut
  {NoStop}%
\bibitem [{\citenamefont {{Drenkhahn}}(2002)}]{Drenkhahn02}%
  \BibitemOpen
  \bibfield  {author} {\bibinfo {author} {\bibfnamefont {G.}~\bibnamefont
  {{Drenkhahn}}},\ }\href {\doibase 10.1051/0004-6361:20020390} {\bibfield
  {journal} {\bibinfo  {journal} {\aap}\ }\textbf {\bibinfo {volume} {387}},\
  \bibinfo {pages} {714} (\bibinfo {year} {2002})},\ \Eprint
  {http://arxiv.org/abs/arXiv:astro-ph/0112509} {arXiv:astro-ph/0112509}
  \BibitemShut {NoStop}%
\bibitem [{\citenamefont {{Granot}}(2011)}]{Granot11bmagjet}%
  \BibitemOpen
  \bibfield  {author} {\bibinfo {author} {\bibfnamefont {J.}~\bibnamefont
  {{Granot}}},\ }\href@noop {} {\bibfield  {journal} {\bibinfo  {journal}
  {ArXiv e-prints}\ } (\bibinfo {year} {2011})},\ \Eprint
  {http://arxiv.org/abs/1109.5315} {arXiv:1109.5315 [astro-ph.HE]} \BibitemShut
  {NoStop}%
\bibitem [{\citenamefont {{Metzger}}\ \emph
  {et~al.}(2011{\natexlab{b}})\citenamefont {{Metzger}}, \citenamefont
  {{Giannios}},\ and\ \citenamefont {{Horiuchi}}}]{Metzger+11magcr}%
  \BibitemOpen
  \bibfield  {author} {\bibinfo {author} {\bibfnamefont {B.~D.}\ \bibnamefont
  {{Metzger}}}, \bibinfo {author} {\bibfnamefont {D.}~\bibnamefont
  {{Giannios}}}, \ and\ \bibinfo {author} {\bibfnamefont {S.}~\bibnamefont
  {{Horiuchi}}},\ }\href {\doibase 10.1111/j.1365-2966.2011.18873.x} {\bibfield
   {journal} {\bibinfo  {journal} {\mnras}\ }\textbf {\bibinfo {volume}
  {415}},\ \bibinfo {pages} {2495} (\bibinfo {year} {2011}{\natexlab{b}})},\
  \Eprint {http://arxiv.org/abs/1101.4019} {arXiv:1101.4019 [astro-ph.HE]}
  \BibitemShut {NoStop}%
\bibitem [{\citenamefont {{Zhang}}\ \emph {et~al.}(2003)\citenamefont
  {{Zhang}}, \citenamefont {{Woosley}},\ and\ \citenamefont
  {{MacFadyen}}}]{Zhang+03jetnum}%
  \BibitemOpen
  \bibfield  {author} {\bibinfo {author} {\bibfnamefont {W.}~\bibnamefont
  {{Zhang}}}, \bibinfo {author} {\bibfnamefont {S.~E.}\ \bibnamefont
  {{Woosley}}}, \ and\ \bibinfo {author} {\bibfnamefont {A.~I.}\ \bibnamefont
  {{MacFadyen}}},\ }\href {\doibase 10.1086/367609} {\bibfield  {journal}
  {\bibinfo  {journal} {\apj}\ }\textbf {\bibinfo {volume} {586}},\ \bibinfo
  {pages} {356} (\bibinfo {year} {2003})},\ \Eprint
  {http://arxiv.org/abs/arXiv:astro-ph/0207436} {arXiv:astro-ph/0207436}
  \BibitemShut {NoStop}%
\bibitem [{\citenamefont {{Tchekhovskoy}}\ \emph {et~al.}(2008)\citenamefont
  {{Tchekhovskoy}}, \citenamefont {{McKinney}},\ and\ \citenamefont
  {{Narayan}}}]{Tchekhovskoy+08grb}%
  \BibitemOpen
  \bibfield  {author} {\bibinfo {author} {\bibfnamefont {A.}~\bibnamefont
  {{Tchekhovskoy}}}, \bibinfo {author} {\bibfnamefont {J.~C.}\ \bibnamefont
  {{McKinney}}}, \ and\ \bibinfo {author} {\bibfnamefont {R.}~\bibnamefont
  {{Narayan}}},\ }\href {\doibase 10.1111/j.1365-2966.2008.13425.x} {\bibfield
  {journal} {\bibinfo  {journal} {\mnras}\ }\textbf {\bibinfo {volume} {388}},\
  \bibinfo {pages} {551} (\bibinfo {year} {2008})},\ \Eprint
  {http://arxiv.org/abs/0803.3807} {arXiv:0803.3807} \BibitemShut {NoStop}%
\bibitem [{\citenamefont {{Kamae}}\ \emph {et~al.}(2006)\citenamefont
  {{Kamae}}, \citenamefont {{Karlsson}}, \citenamefont {{Mizuno}},
  \citenamefont {{Abe}},\ and\ \citenamefont {{Koi}}}]{Kamae06code}%
  \BibitemOpen
  \bibfield  {author} {\bibinfo {author} {\bibfnamefont {T.}~\bibnamefont
  {{Kamae}}}, \bibinfo {author} {\bibfnamefont {N.}~\bibnamefont {{Karlsson}}},
  \bibinfo {author} {\bibfnamefont {T.}~\bibnamefont {{Mizuno}}}, \bibinfo
  {author} {\bibfnamefont {T.}~\bibnamefont {{Abe}}}, \ and\ \bibinfo {author}
  {\bibfnamefont {T.}~\bibnamefont {{Koi}}},\ }\href {\doibase 10.1086/505189}
  {\bibfield  {journal} {\bibinfo  {journal} {\apj}\ }\textbf {\bibinfo
  {volume} {647}},\ \bibinfo {pages} {692} (\bibinfo {year} {2006})},\ \Eprint
  {http://arxiv.org/abs/arXiv:astro-ph/0605581} {arXiv:astro-ph/0605581}
  \BibitemShut {NoStop}%
\bibitem [{\citenamefont {{Hillas}}(1979)}]{Hillas79ICRC}%
  \BibitemOpen
  \bibfield  {author} {\bibinfo {author} {\bibfnamefont {A.~M.}\ \bibnamefont
  {{Hillas}}},\ }in\ \href@noop {} {\emph {\bibinfo {booktitle} {International
  Cosmic Ray Conference}}},\ \bibinfo {series} {International Cosmic Ray
  Conference}, Vol.~\bibinfo {volume} {6}\ (\bibinfo {year} {1979})\
  p.~\bibinfo {pages} {13}\BibitemShut {NoStop}%
\bibitem [{\citenamefont {{Marscher}}\ \emph {et~al.}(1980)\citenamefont
  {{Marscher}}, \citenamefont {{Vestrand}},\ and\ \citenamefont
  {{Scott}}}]{Marscher+80}%
  \BibitemOpen
  \bibfield  {author} {\bibinfo {author} {\bibfnamefont {A.~P.}\ \bibnamefont
  {{Marscher}}}, \bibinfo {author} {\bibfnamefont {W.~T.}\ \bibnamefont
  {{Vestrand}}}, \ and\ \bibinfo {author} {\bibfnamefont {J.~S.}\ \bibnamefont
  {{Scott}}},\ }\href {\doibase 10.1086/158433} {\bibfield  {journal} {\bibinfo
   {journal} {\apj}\ }\textbf {\bibinfo {volume} {241}},\ \bibinfo {pages}
  {1166} (\bibinfo {year} {1980})}\BibitemShut {NoStop}%
\bibitem [{\citenamefont {{Paciesas}}\ \emph {et~al.}(1999)\citenamefont
  {{Paciesas}}, \citenamefont {{Meegan}}, \citenamefont {{Pendleton}},
  \citenamefont {{Briggs}}, \citenamefont {{Kouveliotou}}, \citenamefont
  {{Koshut}}, \citenamefont {{Lestrade}}, \citenamefont {{McCollough}},
  \citenamefont {{Brainerd}}, \citenamefont {{Hakkila}}, \citenamefont
  {{Henze}}, \citenamefont {{Preece}}, \citenamefont {{Connaughton}},
  \citenamefont {{Kippen}}, \citenamefont {{Mallozzi}}, \citenamefont
  {{Fishman}}, \citenamefont {{Richardson}},\ and\ \citenamefont
  {{Sahi}}}]{Paciesas+99}%
  \BibitemOpen
  \bibfield  {author} {\bibinfo {author} {\bibfnamefont {W.~S.}\ \bibnamefont
  {{Paciesas}}}, \bibinfo {author} {\bibfnamefont {C.~A.}\ \bibnamefont
  {{Meegan}}}, \bibinfo {author} {\bibfnamefont {G.~N.}\ \bibnamefont
  {{Pendleton}}}, \bibinfo {author} {\bibfnamefont {M.~S.}\ \bibnamefont
  {{Briggs}}}, \bibinfo {author} {\bibfnamefont {C.}~\bibnamefont
  {{Kouveliotou}}}, \bibinfo {author} {\bibfnamefont {T.~M.}\ \bibnamefont
  {{Koshut}}}, \bibinfo {author} {\bibfnamefont {J.~P.}\ \bibnamefont
  {{Lestrade}}}, \bibinfo {author} {\bibfnamefont {M.~L.}\ \bibnamefont
  {{McCollough}}}, \bibinfo {author} {\bibfnamefont {J.~J.}\ \bibnamefont
  {{Brainerd}}}, \bibinfo {author} {\bibfnamefont {J.}~\bibnamefont
  {{Hakkila}}}, \bibinfo {author} {\bibfnamefont {W.}~\bibnamefont {{Henze}}},
  \bibinfo {author} {\bibfnamefont {R.~D.}\ \bibnamefont {{Preece}}}, \bibinfo
  {author} {\bibfnamefont {V.}~\bibnamefont {{Connaughton}}}, \bibinfo {author}
  {\bibfnamefont {R.~M.}\ \bibnamefont {{Kippen}}}, \bibinfo {author}
  {\bibfnamefont {R.~S.}\ \bibnamefont {{Mallozzi}}}, \bibinfo {author}
  {\bibfnamefont {G.~J.}\ \bibnamefont {{Fishman}}}, \bibinfo {author}
  {\bibfnamefont {G.~A.}\ \bibnamefont {{Richardson}}}, \ and\ \bibinfo
  {author} {\bibfnamefont {M.}~\bibnamefont {{Sahi}}},\ }\href {\doibase
  10.1086/313224} {\bibfield  {journal} {\bibinfo  {journal} {\apjs}\ }\textbf
  {\bibinfo {volume} {122}},\ \bibinfo {pages} {465} (\bibinfo {year}
  {1999})},\ \Eprint {http://arxiv.org/abs/arXiv:astro-ph/9903205}
  {arXiv:astro-ph/9903205} \BibitemShut {NoStop}%
\bibitem [{\citenamefont {{Murase}}(2007)}]{Murase08revisited}%
  \BibitemOpen
  \bibfield  {author} {\bibinfo {author} {\bibfnamefont {K.}~\bibnamefont
  {{Murase}}},\ }\href {\doibase 10.1103/PhysRevD.76.123001} {\bibfield
  {journal} {\bibinfo  {journal} {\prd}\ }\textbf {\bibinfo {volume} {76}},\
  \bibinfo {eid} {123001} (\bibinfo {year} {2007})},\ \Eprint
  {http://arxiv.org/abs/0707.1140} {arXiv:0707.1140} \BibitemShut {NoStop}%
\bibitem [{\citenamefont {{Wanderman}}\ and\ \citenamefont
  {{Piran}}(2010)}]{Wanderman+10grbsfr}%
  \BibitemOpen
  \bibfield  {author} {\bibinfo {author} {\bibfnamefont {D.}~\bibnamefont
  {{Wanderman}}}\ and\ \bibinfo {author} {\bibfnamefont {T.}~\bibnamefont
  {{Piran}}},\ }\href {\doibase 10.1111/j.1365-2966.2010.16787.x} {\bibfield
  {journal} {\bibinfo  {journal} {\mnras}\ }\textbf {\bibinfo {volume} {406}},\
  \bibinfo {pages} {1944} (\bibinfo {year} {2010})},\ \Eprint
  {http://arxiv.org/abs/0912.0709} {arXiv:0912.0709 [astro-ph.HE]} \BibitemShut
  {NoStop}%
\bibitem [{\citenamefont {{Gehrels}}\ \emph {et~al.}(2009)\citenamefont
  {{Gehrels}}, \citenamefont {{Ramirez-Ruiz}},\ and\ \citenamefont
  {{Fox}}}]{Gehrels09rev}%
  \BibitemOpen
  \bibfield  {author} {\bibinfo {author} {\bibfnamefont {N.}~\bibnamefont
  {{Gehrels}}}, \bibinfo {author} {\bibfnamefont {E.}~\bibnamefont
  {{Ramirez-Ruiz}}}, \ and\ \bibinfo {author} {\bibfnamefont {D.~B.}\
  \bibnamefont {{Fox}}},\ }\href {\doibase
  10.1146/annurev.astro.46.060407.145147} {\bibfield  {journal} {\bibinfo
  {journal} {\araa}\ }\textbf {\bibinfo {volume} {47}},\ \bibinfo {pages} {567}
  (\bibinfo {year} {2009})},\ \Eprint {http://arxiv.org/abs/0909.1531}
  {arXiv:0909.1531 [astro-ph.HE]} \BibitemShut {NoStop}%
\bibitem [{\citenamefont {{Greiner}}(2011)}]{GreinerWEBPAGE}%
  \BibitemOpen
  \bibfield  {author} {\bibinfo {author} {\bibfnamefont {J.}~\bibnamefont
  {{Greiner}}},\ }\href@noop {} {\bibfield  {journal} {\bibinfo  {journal}
  {http://www.mpe.mpg.de/~jcg/grbgen.html}\ } (\bibinfo {year}
  {2011})}\BibitemShut {NoStop}%
\end{thebibliography}%

\newpage

\end{document}